\documentclass[showpacs,twocolumn,superscriptaddress,prl,preprintnumbers,nofootinbib]{revtex4}
\usepackage{amsmath,amssymb,bm,graphicx,color,gensymb}

\begin{document}

\title{Disorder-Induced Mimicry of a Spin Liquid  in YbMgGaO$_4$}

\author{Zhenyue Zhu}
\affiliation{Department of Physics and Astronomy, University of California, Irvine, California
92697, USA}
\author{P. A. Maksimov}
\affiliation{Department of Physics and Astronomy, University of California, Irvine, California
92697, USA}
\author{Steven R. White}
\affiliation{Department of Physics and Astronomy, University of California, Irvine, California
92697, USA}
\author{A. L. Chernyshev}
\affiliation{Department of Physics and Astronomy, University of California, Irvine, California
92697, USA}
\date{\today}
\begin{abstract}
We suggest that a randomization of the pseudo-dipolar interaction in the spin-orbit-generated 
low-energy Hamiltonian of YbMgGaO$_4$ due to an inhomogeneous charge environment from a natural 
mixing of Mg$^{2+}$ and Ga$^{3+}$ can give rise to 
orientational spin disorder and mimic a spin-liquid-like state.
In the absence of such quenched disorder,  $1/S$ and  density matrix renormalization group 
calculations both show robust ordered states for the physically relevant phases of the model.
Our scenario is consistent with the available experimental data and further experiments
are proposed to support it.
\end{abstract}
\maketitle

Dating back to Wannier's pioneering study of the Ising model \cite{wannier}, 
triangular lattice models and materials with frustrating antiferromagnetic interactions 
have served as fertile playgrounds for new ideas
\cite{spinliquid,Huse88,Oguchi83,Miyake92,Chubukov94,Leung93,Lhuillier94,capriotti99,white07}.
These systems continue to draw significant experimental
\cite{Collins97,Nakatsuji05,Olariu06,expBaCoSbO1,new_MM} 
and theoretical interest because they exhibit many intriguing  
novel ordered  states 
\cite{kawamura,korshunov,chubukov91,yamamoto,starykh14,eggert,starykh_review} and
unusual continuum-like spectral features  
\cite{Coldea01,zheng06,starykh10,expBaCoSbO2016,JeGeun,RMP,starykh06,triangle,triSqw} 
and especially because they provide a setting for spin-liquid states 
\cite{Mishmash13,ZhuWhite,Yigbal,Chubukov,Jolicquer,Gazza,Bishop,Kaneko,Evertz,MoessnerSondhi01,SavaryBalentsSL16}. 

Among the latest experimental discoveries \cite{expBaCoSbO1,new_MM},
a rare-earth triangular-lattice  antiferromagnet YbMgGaO$_4$ has recently emerged 
as a new candidate for a quantum spin liquid of the effective spin-$1/2$ degrees of freedom of Yb$^{3+}$ ions
\cite{SciRep,Chen1}.  
It has been argued  that the spin-orbit origin of its magnetic properties 
and the pseudo-spin nature of the low-energy states 
with highly anisotropic effective spin interactions may
potentially open a new route to realizing quantum spin liquids \cite{Chen1,Chen2,muons}.
While the lack of ordering, anomalous specific heat, and especially continuum-like excitations in inelastic 
neutron scattering \cite{Chen2,MM} all provide strong support to the idea of an intrinsic spin liquid, 
other experimental findings are increasingly at odds with this picture.

First, in magnetization vs field measurements, there is no sharpening of the transition to the saturated phase 
upon lowering the temperature, and the lack of the upward curvature in $M(H)$ at the lowest  $T$'s \cite{SciRep,Chen1} 
is indicative of low quantum fluctuations in the ground state \cite{Griffits64}. 
Second,   in the high-field polarized phase, neutron scattering shows that continuum-like excitations persist,
with significant smearing of magnon lines that are expected to be sharp \cite{MM}.
In addition, an apparent absence of any detectable contribution of spin excitations to  
thermal conductivity  down to the lowest temperatures, 
accompanied by a strong deviation of the phonon part from the ballistic $T^3$ form \cite{kappa}, 
both  suggest strong scattering effects.
These, combined with the anomalously broadened higher-energy Yb$^{3+}$ doublet structure \cite{MM,Yuesheng17}
and a ubiquitous mixing of Mg$^{2+}$ and Ga$^{3+}$ ions in the non-magnetic layers \cite{SciRep, MM},
implicate disorder as a key contributor to the observed properties \cite{Yuesheng17}.

In this Letter, we first argue that a hypothetical, disorder-free version of YbMgGaO$_4$
should exhibit a robust collinear/stripe magnetic order.
We demonstrate this by extending  the well-studied phase diagram of the triangular-lattice Heisenberg  
$J_1\!-\!J_2$ model, 
which is known to have an extensive spin-liquid region for $S\!=\!1/2$
\cite{ZhuWhite,Yigbal,Chubukov,Jolicquer,Gazza,Bishop,Kaneko,Evertz},
to the anisotropic version of the model that corresponds to the types of 
anisotropy allowed in YbMgGaO$_4$ with realistic restrictions from experiments.
A significant $XXZ$ anisotropy present in YbMgGaO$_4$ 
 suppresses the spin-liquid region of the phase diagram,
and the pseudo-dipolar interactions further diminish it. 
Both types of anisotropy lower the symmetry and produce gaps in the excitation spectra, 
reducing quantum fluctuations that suppress the ordered states. 

We then suggest that the stripe order is fragile to an orientational disorder 
that can be easily produced via a randomization of the subleading pseudo-dipolar  interactions. 
The physical reason of such a sensitivity is a small energetic barrier, $\delta E\!\sim\! 0.03 J_1$ per site, 
between the stripe phases of different spatial orientations, which,  in the absence of the pseudo-dipolar terms, 
are selected by   order-by-disorder fluctuations. 
Thus, we propose that the spin-liquid-like state in  YbMgGaO$_4$ is disorder induced and is composed 
of nearly classical, orientationally randomized, short-range stripe-like spin domains. The 
quenched,  spatially-fluctuating charge environment of the magnetic Yb$^{3+}$ ions
due to random site occupancies of Mg$^{2+}$ and Ga$^{3+}$ ions is seen as a likely culprit,
affecting the low-energy effective spin Hamiltonian through the spin-orbit coupling.

\emph{Model.}---%
Although the magnetism of YbMgGaO$_4$ is dominated by   spin-orbit coupling, which 
can  result in large spin anisotropies of various types
\cite{Shannon16,Ross11,Savary12,MZh12}, it is restricted by the high symmetry of the 
lattice  \cite{Chen1,Chen2}, yielding the familiar $XXZ$ anisotropy accompanied 
by the so-called pseudo-dipolar terms.
Moreover, the local character of the $f$ shells on Yb dictates that the dominant interactions are 
between the nearest-neighbor spins,  further restricting possible spin models.

\begin{figure}[t]
\includegraphics[width=\linewidth]{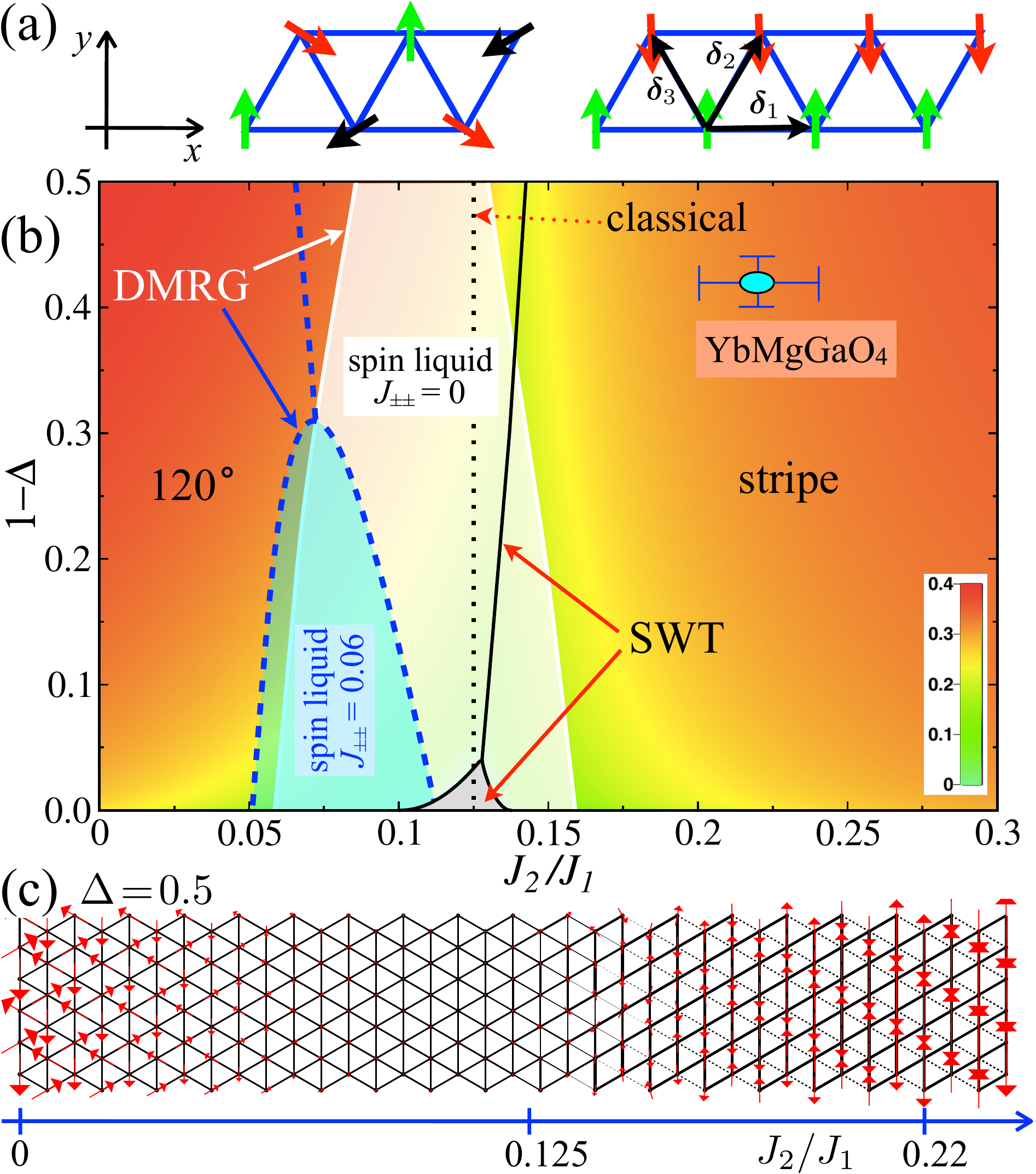}
\vskip -0.2cm
\caption{(a) Axes, primitive vectors,  and a sketch of the $120{\degree}$ and stripe states. (b)
$1\!-\!\Delta$ vs $J_2$ phase diagram of the $XXZ$ model (\ref{HXXZ}). 
The $\langle S\rangle$ color map and boundaries (solid lines) are by the SWT; the dotted line is 
the classical phase boundary. 
The shaded white area is the spin-liquid region by the DMRG;  see the text. The dashed line with the
shaded region is the same for the  model with ${\cal H}_{\rm pd}$ with  $|J_{\pm\pm}|\!=\!0.06$; see  
Fig.~\ref{Fig_Jpm}. The error bars mark YbMgGaO$_4$ parameters from Ref.~\cite{MM}. 
(c) The DMRG scan of (\ref{HXXZ}) vs $J_2$ for $\Delta\!=\!0.5$ with up to $2000$ states.}
\label{Fig1}
\vskip -0.4cm
\end{figure}

Thus, we are compelled to explore the phase diagram of the following $S\!=\!1/2$ model as  relevant 
to YbMgGaO$_4$ \cite{SciRep,Chen1,Chen2,MM} 
and also to a broader family of the rare-earth  triangular-lattice materials \cite{Cava16}:
${\cal H}\!=\!{\cal H}_{XXZ}^{J_1-J_2}\!+\!{\cal H}_{\rm pd}$, with
\begin{eqnarray}
{\cal H}_{XXZ}^{J_1-J_2}=\sum_{\langle ij\rangle_{n}}
J_{n}\left(S^{x}_i S^{x}_j+S^{y}_i S^{y}_j+\Delta S^{z}_i S^{z}_j\right),
\label{HXXZ}
\end{eqnarray}
\vskip -0.15cm \noindent
where the sums are over the (next-)nearest neighbors with $J_{1}\!>\!J_{2}\!\geq\!0$, the $XXZ$ anisotropy 
$0\!\leq\!\Delta\!\leq\!1$, and the pseudo-dipolar terms  introduced as 
\cite{MM,Chen1,Chen2}
\begin{eqnarray}
{\cal H}_{\rm pd}=J_{\pm\pm}\sum_{\langle ij\rangle}\left(e^{i\tilde{\varphi}_\alpha} S^{+}_i S^{+}_j
+e^{-i\tilde{\varphi}_\alpha} S^{-}_i S^{-}_j\right),
\label{HJpm}
\end{eqnarray}
\vskip -0.15cm \noindent
where $S^{\pm}\!=\!S^x\pm iS^y$ and
$\tilde{\varphi}_\alpha\!=\!\{0,-2\pi/3,2\pi/3\}$ are the bond-dependent phases for the 
primitive vectors ${\bm \delta}_\alpha$, with ${\bm \delta}_\alpha$'s and $x$ and $y$ axes as in 
Fig.~\ref{Fig1}(a). Although this  is not obvious 
from (\ref{HJpm}) \cite{footnote1}, the pseudo-dipolar terms favor the direction of the spins on a bond to be either parallel
or perpendicular to the bond \cite{MZh12}. 
Because of the high symmetry of the lattice, the Dzyaloshinsky-Moriya interactions are 
forbidden \cite{SciRep,Chen3} and we also omit the  
 couplings of  $S^{x(y)}$'s to the out-of-plane $S^{z}$'s, referred to as 
the $J_{z\pm}$ terms, as they are negligible in YbMgGaO$_4$  \cite{MM,Chen1} and do not 
affect our conclusions.
An intuitive derivation of the  Hamiltonian  
is given in \cite{supp}.

\emph{XXZ only.}---%
In YbMgGaO$_4$, electron spin resonance (ESR), magnetic susceptibility, and neutron scattering \cite{Chen1,MM}
 have suggested strong $XXZ$ anisotropy, $\Delta\!\sim\!0.5$, and  
put rather stringent bounds on the pseudo-dipolar terms, indicating their 
subleading role. Thus, we study the pure $XXZ$ model (\ref{HXXZ}) first,
considering effects of the pseudo-dipolar terms next.
The anisotropy for $J_1$ and $J_2$ bonds, $\Delta_1$ and $\Delta_2$, is assumed equal \cite{MM}, 
as it originates from the magnetic state of Yb$^{3+}$
ions, with no qualitative changes expected for $\Delta_1\!\neq\!\Delta_2$.

While the Heisenberg version of (\ref{HXXZ}) at $\Delta\!=\!1$ is well explored 
\cite{ZhuWhite,Yigbal,Chubukov,Jolicquer,Gazza,Bishop,Kaneko,Evertz},
its  anisotropic extension has been studied only rarely \cite{Ivanov,Pires}.
For $J_2/J_1\!<\!1$,  two ordered states  compete,  the $120{\degree}$
and the collinear state, where in the latter ferromagnetic rows (``stripes'') of
spins align antiferromagnetically; see Fig.~\ref{Fig1}(a). 
Their classical  energies are
$E^{120{\degree}}_{\rm gs}\!\!=\!-3(J_1/2-J_2)$ and  
$E^{\rm str}_{\rm gs}\!=\!-J_1-J_2$ (per $NS^2$),  
yielding a transition  at $J_2\!=\!J_1/8$ \cite{Chubukov,Jolicquer} independent  of $\Delta$.
It is important to note that   $XXZ$
anisotropy leads to an overlap of the $J_2$ ranges of stability for 
magnon spectra of the competing phases \cite{Ivanov,supp}. 
This implies that the spin-wave instabilities do not yield an intermediate magnetically disordered 
state for $S\!\gg\!1$, favoring instead a direct transition between the two orders. 

The $J_2\!-\!\Delta$ phase diagram of ${\cal H}_{XXZ}^{J_1-J_2}$  
for $S\!=\!1/2$, obtained via the  spin-wave theory (SWT) and density matrix renormalization group (DMRG) calculations,
is shown in Fig.~\ref{Fig1}(b).
The color map shows the ordered moment $\langle S\rangle$ and the $\langle S\rangle\!=\!0$ boundaries 
of a non-magnetic phase  (gray)  according to the SWT. 
The solid black line marks the crossing of $\langle S\rangle$ from the $120{\degree}$ to the stripe phase.
It outlines a region where the SWT predicts a direct transition with no intermediate state. 
Note that the SWT ground state energies indicate this transition to be on the left of the classical $J_2\!=\!J_1/8$ line 
for $\Delta\!<\!1$ \cite{supp}.

Figure~\ref{Fig1}(c) shows a DMRG calculation of the model (\ref{HXXZ}) for $\Delta\!=\!0.5$
where $J_2$ is varied along the length of the cylinder so that different phases appear at different regions.
The orders are pinned at the boundaries and the spin patterns give a faithful visual extent of their phases.  
Similar scans for several $\Delta$'s allow us to map out  the phase diagram of the model
\cite{ZhuWhite,hexZhuWhite}. 
To roughly estimate the $J_2$ boundaries for the spin liquid (SL), we use the cutoff value of $\langle S\rangle\!=\!0.05$,
below which the system is assumed to be in a SL state.
This procedure matches the SL boundaries for the isotropic ($\Delta\!=\!1$) $J_1\!-\!J_2$ model found in 
Ref.~\cite{ZhuWhite}
by a more accurate method.  
The resultant extent of the SL phase is shown in Fig.~\ref{Fig1}(b) by the white shaded area.
We note that the $\langle S\rangle$ cutoff value that we use  may overestimate the SL region at $\Delta\!<\!1$, as 
the anisotropy tends to stabilize ordered phases, while the SWT clearly underestimates it, as expected.

The ellipse with  
error bars in Fig.~\ref{Fig1}(b) marks $J_2/J_1\!=\!0.22(2)$ and $\Delta\!=\!0.58(2)$, 
proposed for YbMgGaO$_4$  \cite{MM}.   For these parameters (with $J_{\pm\pm}\!=\!0$), we find
a close agreement between the DMRG  and SWT on the ordered moment, 0.29  and 0.32, respectively,
implying that YbMgGaO$_4$ is deep in the stripe phase.

\begin{figure}[t]
\includegraphics[width=\linewidth]{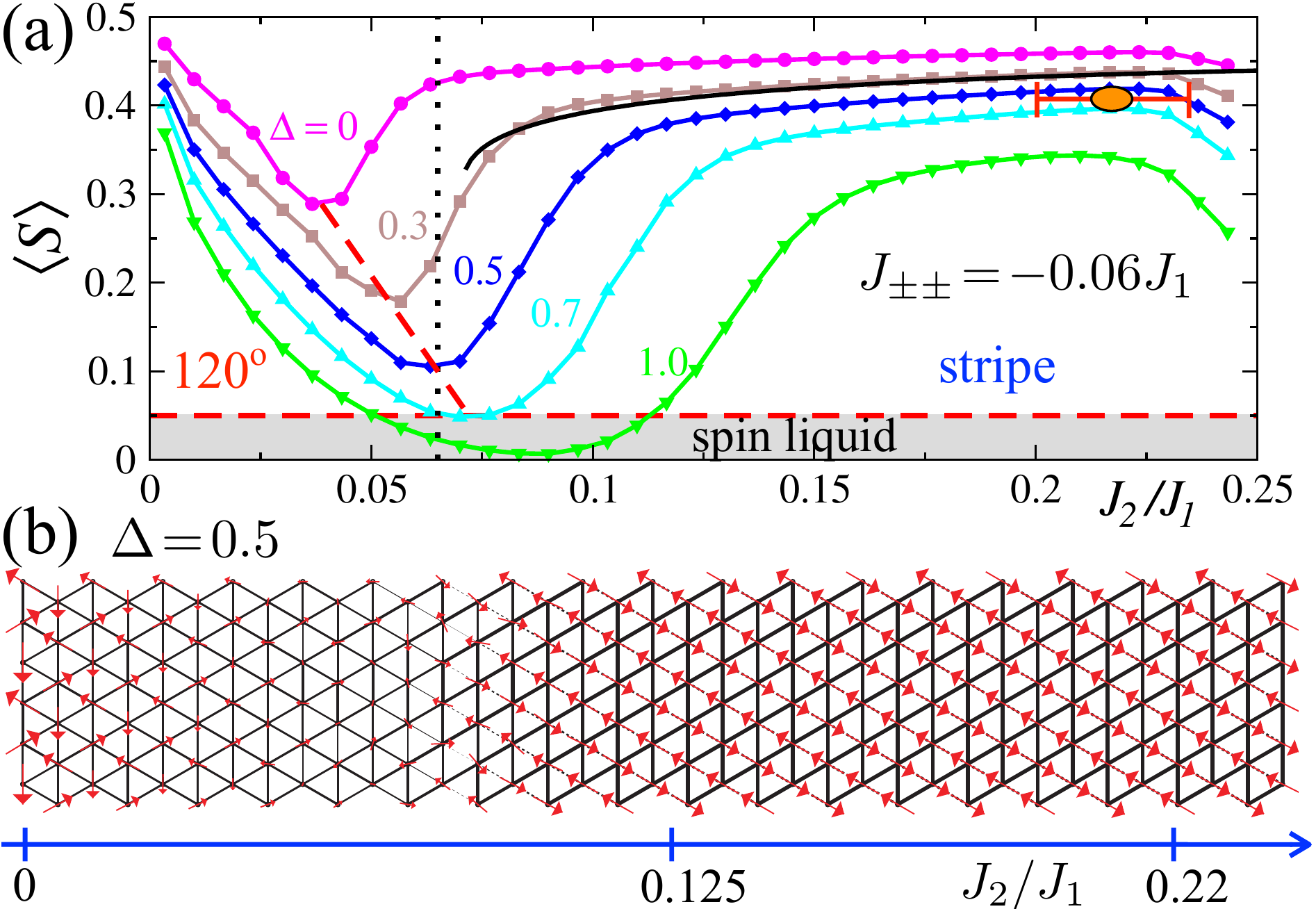}
\vskip -0.2cm
\caption{(a) 
    DMRG results for $\langle S\rangle$ vs $J_2$  with $|J_{\pm\pm}|\!=\!0.06J_1$.
Dotted and dashed lines denote classical and DMRG phase boundaries, respectively. 
The error bar is the same as in Fig.~\ref{Fig1}(b).  The solid black line is the SWT result 
for $\Delta\!=\!0.5$. (b) A long-cylinder DMRG scan for $\Delta\!=\!0.5$ and 
$J_{\pm\pm}\!=\!-0.06J_1$.}
\label{Fig_Jpm}
\vskip -0.4cm
\end{figure}

\emph{Pseudo-dipolar terms.}---%
The anisotropic terms in (\ref{HJpm}) explicitly break the  
U(1) symmetry of the $XXZ$ model (\ref{HXXZ})
and are expected to pin the  spin directions to the lattice.  
This is indeed true for the stripe phase, in which the pseudo-dipolar terms
make the spin orientation   parallel ($J_{\pm\pm}\!<\!0$) or perpendicular ($J_{\pm\pm}\!>\!0$) 
to the stripe direction  \cite{supp} as in Figs.~\ref{Fig_Jpm}(b) or \ref{Fig1}(a); see also \cite{Chen3}.
From the $1/S$ perspective, no pinning and no change of the classical energy occurs  due to (\ref{HJpm}) for the 
$120{\degree}$ phase, which, however, remains stable \cite{supp}. 
On the other hand, the partially frustrated  pseudo-dipolar  terms in (\ref{HJpm}) 
lower the classical energy of the stripe phase by $-4|J_{\pm\pm}| S^2N$ and expand its stability range by shifting 
the classical phase boundary to a lower $J_2\!=\!J_1/8\!-\!|J_{\pm\pm}|$.

In Figs.~\ref{Fig_Jpm} and \ref{Fig1}(b), 
we show the effect of adding $J_{\pm\pm}$ to the model, using
$|J_{\pm\pm}|\!=\!0.06J_1$, as suggested by 
ESR \cite{Chen1}.
The classical transition between the $120{\degree}$ and stripe phases is at $J_2\!=\!0.065J_1$ for this value of 
$|J_{\pm\pm}|$, with the DMRG long-cylinder scans showing it tilting toward smaller $J_2$ at smaller $\Delta$. 
Using the same generous criteria for the spin liquid as above, the DMRG results show that 
$J_{\pm\pm}$ shrinks the SL region [light blue in Fig.~\ref{Fig1}(b)], 
and moves it farther from the YbMgGaO$_4$ parameters.
It also strengthens the stripe order [Fig.~\ref{Fig_Jpm}(a)],
in   close agreement with the SWT (solid line). The agreement for the ordered moment 
for YbMgGaO$_4$ parameters \cite{MM}  is very close, $\langle S\rangle\!\approx \! 0.419$ 
$(0.433)$ by DMRG (SWT), and the magnitude of the order parameter is large.

Thus, in this model for YbMgGaO$_4$, the easy-plane and pseudo-dipolar anisotropies  both
lead to a stronger stripe order. Yet, the 
experiments show no sign of it.

 Alternative sets of parameters with much larger 
values  of $|J_{\pm\pm}|\!=\!0.26J_1$ \cite{Chen4}  and $0.69J_1$ \cite{MM}
were obtained  by fitting the high-field magnon dispersion  in YbMgGaO$_4$ 
\cite{MM} without  the $J_2$ term in (\ref{HXXZ}). 
Both values strongly deviate from the ESR data \cite{Chen1} and imply an almost classical stripe state with 
nearly saturated ordered moments and large magnon gaps \cite{supp}, inconsistent with 
the observed substantial spectral weight at low energies \cite{MM}.  
For  $|J_{\pm\pm}|\!\agt\!0.2$, there is no $120{\degree}$ state left  in the phase diagram to compete with, 
leaving no  SL state in sight.

\begin{figure}[t]
\includegraphics[width=\linewidth]{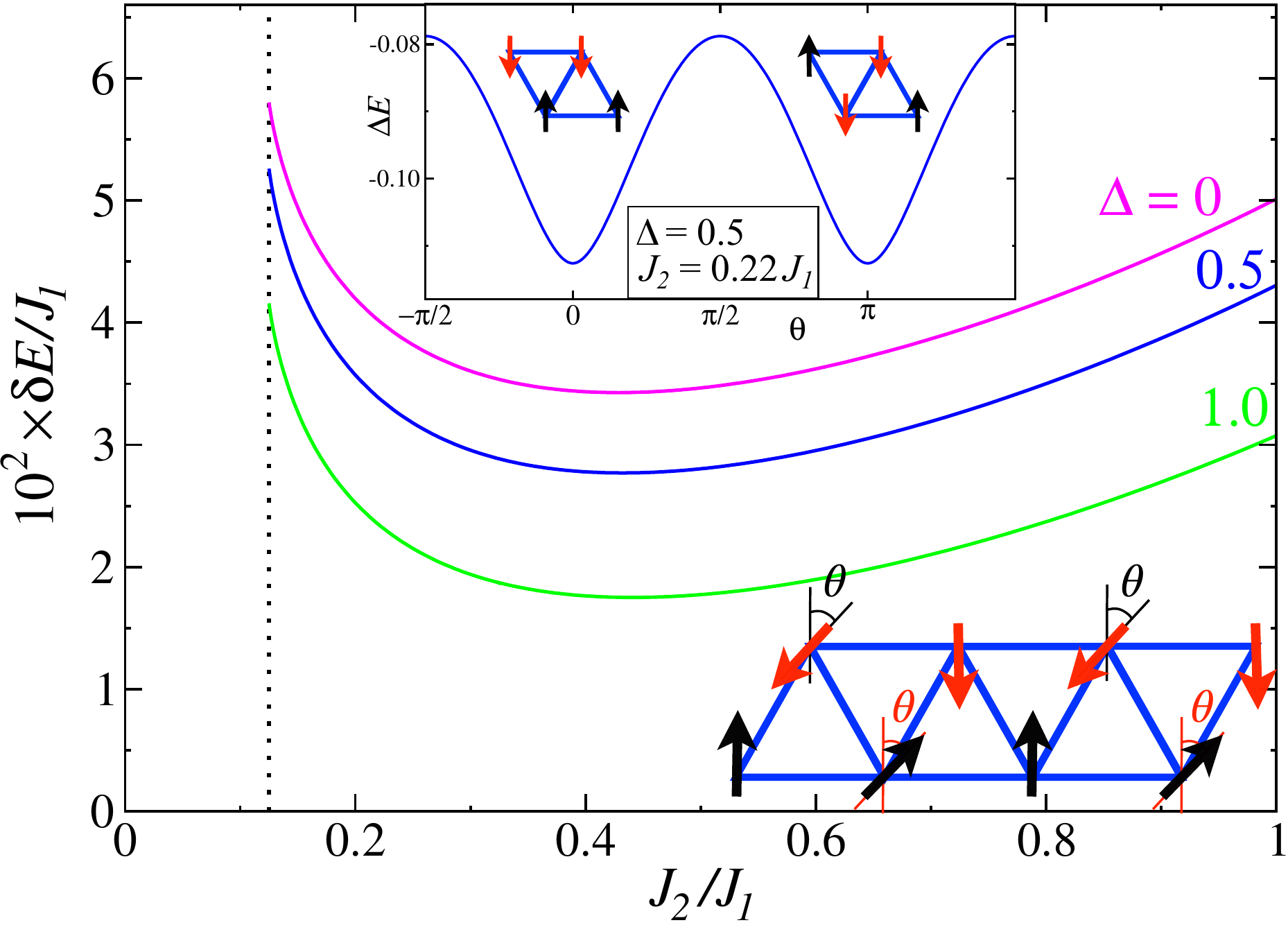}
\vskip -0.3cm
\caption{Energy barrier between the stripe states of different orientations in the $XXZ$  
$J_1\!-\!J_2$ model vs $J_2$ for various $\Delta$ and $S\!=\!1/2$. 
Upper inset: Quantum energy correction vs  angle $\theta$. 
Lower inset: A sketch of the degenerate classical ground states with 
$\theta\!=\!0(\pi)$ corresponding to two stripe orientations. }
\label{barrier}
\vskip -0.5cm
\end{figure}

\emph{Barrier.}---%
Before we attempt to reconcile our finding of strong stripe order in the model with the lack of order in YbMgGaO$_4$,
we give the $J_1\!-\!J_2$ $XXZ$ model (\ref{HXXZ}) a second look.
Classically, in the absence of the pseudo-dipolar terms, the stripe phases of Fig.~\ref{Fig1}(a) 
are degenerate with a manifold of  spiral phases in Fig.~\ref{barrier}, in which four spins in the two side-sharing 
triangles add up to zero \cite{Jolicquer,Chubukov}. Their degeneracy is lifted  
via order-by-disorder mechanism \cite{CCL,Henley}, selecting the three stripe states that break rotational lattice symmetry. 
The tunneling barrier between them, $\delta E(J_2,\Delta)/N$, shown in Fig.~\ref{barrier},
is obtained from  
the quantum energy correction
$\Delta E(\theta)\!=\!c+\frac{1}{2}\sum_{\bf k} \varepsilon_{\bf k} (\theta)$, 
where $c\!=\!-(J_1+J_2)S$ and $\varepsilon_{\bf k} (\theta)$ is the magnon energy, 
which depends on the angle $\theta$ of the spiral state 
from the  degenerate classical  manifold.
As one can see from Fig.~\ref{barrier}, the  tunneling barrier is small, $\delta E\!\sim\!0.03J$ per site,
similar to the $J_1\!-\!J_2$ model on the square lattice \cite{Ioanis}. 
Thus,  in the $XXZ$ model, despite being strongly ordered, the stripe phases of different  orientations
are separated by a low energetic barrier. 

\begin{figure}[t]
\includegraphics[width=\linewidth]{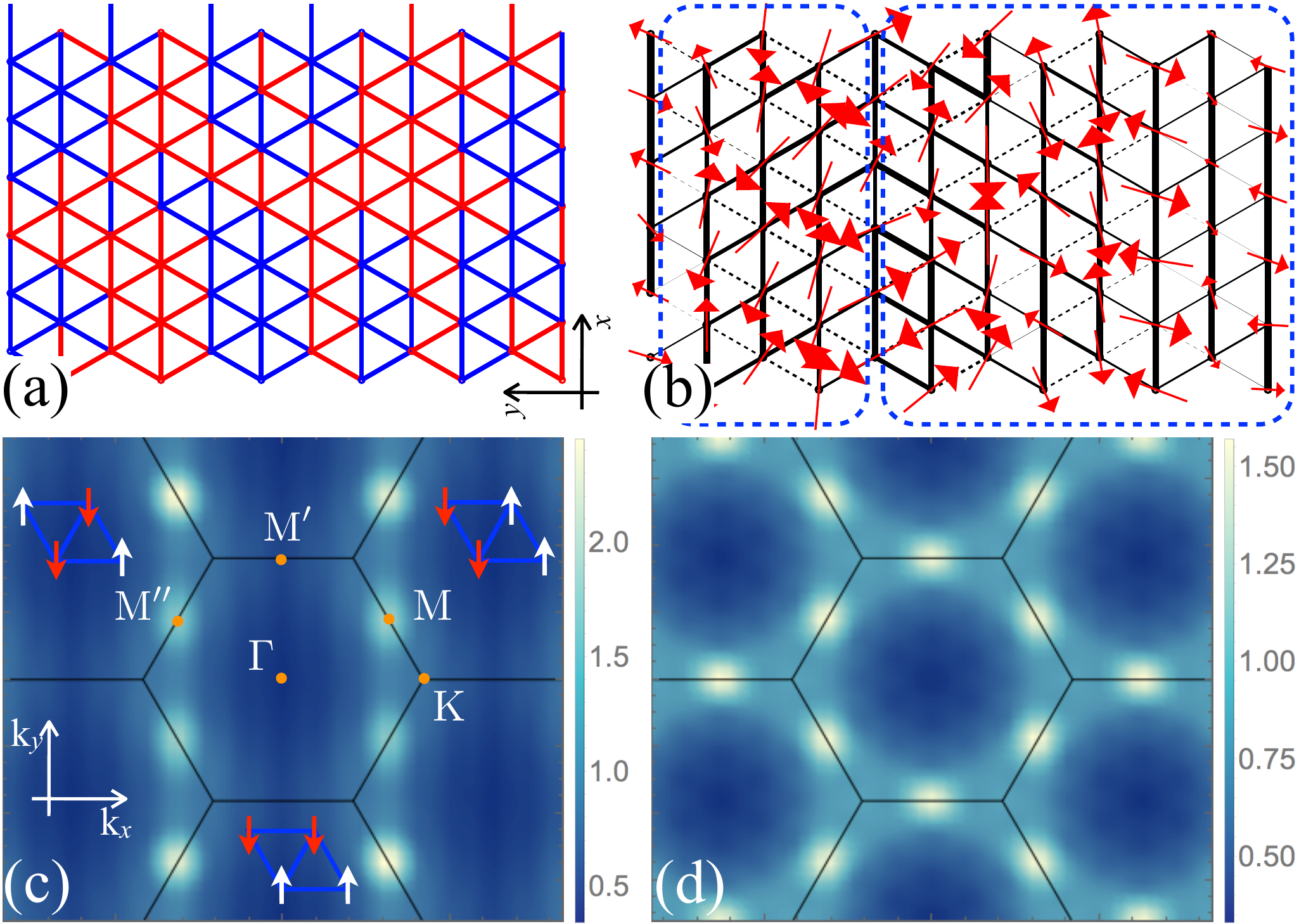}
\vskip -0.2cm
\caption{(a) Positive and negative $J_{\pm\pm}$ bonds in a typical disorder realization. 
(b) Two  stripe domains (dashed boxes) for random $|J_{\pm\pm}|\!=\!0.2J_1$, $\langle S\rangle$ up to 0.33. 
(c) ${\cal S}({\bf q})$ \cite{footnoteSq} 
for random $|J_{\pm\pm}|\!=\!0.1(0.05)J_1$. (d) Averaged ${\cal S}({\bf q})$ from (c); see the text.}
\label{Fig4}
\vskip -0.4cm
\end{figure}

\emph{Disorder.}---%
As   discussed above, a number of experiments indicate a substantial disorder in the 
low-energy effective spin Hamiltonian of YbMgGaO$_4$ \cite{SciRep,MM,kappa,Yuesheng17}. 
Most direct are the neutron studies, 
suggesting strong variations in the effective $g$ factors and, possibly, magnetic couplings \cite{Yuesheng17}
due to a random charge environment from mixing of the non-magnetic Mg$^{2+}$ and Ga$^{3+}$.

We do not attempt to analyze all forms of disorder that can  naturally occur in the Hamiltonian
(\ref{HXXZ}) and (\ref{HJpm}). Instead, we propose that a disorder in the $J_{\pm\pm}$ terms should be 
potentially very destructive. Because of their pseudo-dipolar nature, random $J_{\pm\pm}$'s
are not unlike fluctuating pinning fields that can locally stabilize stripes with 
different orientations by overcoming the low tunneling barrier between them. 
In addition, for the  relevant values of $|J_{\pm\pm}|\!\sim\! 0.1J_1\!\agt\!\delta E$, 
fluctuations of the diagonal elements of the exchange tensor at the level 
of $0.1\!-\!0.2J_1$, that are consistent with the variations suggested  in Ref.~\cite{Yuesheng17}, 
translate into completely random $J_{\pm\pm}$ \cite{supp}.

We have performed DMRG calculations of the $J_1\!-\!J_2$ $XXZ$ model (\ref{HXXZ}) 
with YbMgGaO$_4$ parameters, $\Delta\!=\!0.58$ and $J_2\!=\!0.22J_1$ \cite{MM}, 
and random $J_{\pm\pm}$ (\ref{HJpm}). 
We have used different random disorder realizations,
such as in Fig.~\ref{Fig4}(a), with a binary distribution 
of $J_{\pm\pm}$ of alternating sign and a global constraint 
of the same number of positive and negative $J_{\pm\pm}$ bonds to reduce the finite-size bias.
We used the values of $|J_{\pm\pm}|/J_1\!=\!0.05$, $0.1$, and $0.2$ on the $6\!\times\!12$ cluster.
The results are as follows.

For   large values of  random $|J_{\pm\pm}|\!=\!0.2J_1$, the ground states tend to contain static, visibly 
disordered spin domains with mixed stripe orientations and large   ordered moments;
see Fig.~\ref{Fig4}(b). For   smaller $|J_{\pm\pm}|$, more interesting states appear.  
First, there is no clear real-space order without pinning fields, as  
in a disorder-free U(1)-symmetric $XXZ$ model, yet the structure factor \cite{footnoteSq},  obtained from 
${\cal S}^{\alpha\beta}_{\bf q}\!=\!\sum_{i,j}\langle S_i^\alpha S_j^\beta\rangle e^{i{\bf q}({\bf R}_i-{\bf R}_j)}$,
 shows broadened peaks  at   \emph{two} M points, which  are associated 
with \emph{two different} stripe orderings; see Fig.~\ref{Fig4}(c).
We note that the $6\times 12$ DMRG cluster strongly disfavors the state with stripes along the shorter 
direction of the cylinder, parallel to the open boundaries, that would show itself as a peak at the M$^\prime$ 
points in Fig.~\ref{Fig4}(c). 
 
Upon a careful investigation with  pinning fields, we conclude that the observed state  is  
a \emph{stripe-superposition state}, in which spins continue to fluctuate collectively between the two
stripe states allowed by the cluster. A hint of such a state can also be seen at the right edge in Fig.~\ref{Fig4}(b). 
As opposed to a spin liquid, the degeneracy of such a superposed state is not extensive.
This finding implies that the randomization of $J_{\pm\pm}$ leads to an effective restoration of the 
$Z_3$ lattice symmetry, broken in each individual stripe state.
Whether such  stripe-superposition states will be pinned to form single-stripe domains on a larger length scale,
or they will survive as localized fluctuating states, remains an open question.
   
Note that both $|J_{\pm\pm}|/J_1\!=\!0.05$ and $0.1$ yield nearly identical structure factors,
with the smaller value already sufficient to destroy the long-range stripe order,
supporting our hypothesis on its fragility to an orientational disorder.
To overcome the lack of the third stripe direction in the DMRG cluster and provide a faithful view of a 
response of a spatially isotropic system, we have performed 
an averaging of the structure factor [see Fig.~\ref{Fig4}(d)],
with the results very similar to the ${\cal S}({\bf q})$ in the neutron-scattering data 
for YbMgGaO$_4$ \cite{MM}.

Altogether, the randomization of the small pseudo-dipolar term in the model description of YbMgGaO$_4$ 
results in the disordered stripe ground states that can successfully mimic a spin liquid.  
Further experimental verifications of the proposed picture include possible
freezing at lower temperatures, as the current lowest-temperature measurements \cite{MM} are at 
$T\!\sim\!0.05J_1\!\sim\! |J_{\pm\pm}|$, and the spin pseudo-gap in the dynamical response at low energies 
at the M points as a remnant of the anisotropy-induced gaps 
in the magnon spectra \cite{supp}. 
The proposed scenario implies that the anomalously low $T$ power in the specific heat should emerge as a 
result of disorder.

\emph{Summary.}---%
We have investigated a generalization of the isotropic $J_1\!-\!J_2$ triangular-lattice model, known to support a 
spin-liquid state, and have found that the anisotropic interactions 
significantly diminish the spin-liquid region of the phase diagram. Our analysis finds no additional 
transitions near the experimentally relevant range of parameters, 
putting YbMgGaO$_4$ firmly in the stripe-ordered state.
At the same time, the stripe states are shown to be fragile toward orientational disorder.
The randomization of the pseudo-dipolar interactions due to the spatially fluctuating 
charge environment of the magnetic ions generates a mimicry of a spin-liquid state 
in the form of short-range stripe or stripe-superposition domains.
This scenario is likely to be relevant to other rare-earth-based quantum magnets.

\begin{acknowledgments}
We  thank Andrey Chubukov and Natalia Perkins for fruitful conversations,
Mike Zhitomirsky and Sid Parameswaran for pointed comments, Oleg Starykh for patient explanations and helpful feedback, 
and Kate Ross for an enlightening discussion. We are immensely grateful to Ioanis Rousochatzakis for sharing his old 
notes on the order-by-disorder effect and for many valuable remarks.
We are particularly indebted to Martin Mourigal for his indispensable comments, numerous conversations, 
constructive attitude, and extremely useful insights.
This work was supported by the U.S. Department of Energy,
Office of Science, Basic Energy Sciences under Award No. DE-FG02-04ER46174 (P.  A. M. and A.  L. C.)
and by the NSF through Grant DMR-1505406 (Z. Z. and S.  R. W.).
A.  L. C. thanks Aspen Center for Physics, where part of this work was done. 
The work at Aspen was supported in part by NSF Grant No. PHYS-1066293.
\end{acknowledgments}

{\it Note added.} Recently, we became aware of work that 
supports  our findings \cite{Wang17}.

\vskip -0.3cm \noindent



\newpage \
\newpage
\onecolumngrid
\begin{center}
{\large\bf Disorder-induced mimicry of a spin liquid  in YbMgGaO$_4$:  Supplemental Material}\\ 
\vskip0.35cm
Z. Zhu,$^1$ P. A. Maksimov,$^1$ S. R. White,$^1$ and A. L. Chernyshev$^1$\\
\vskip0.15cm
{\it \small $^1$Department of Physics and Astronomy, University of California, Irvine, California
92697, USA}\\
{\small (Dated: June 9, 2017)}\\
\vskip 0.1cm \
\end{center}
\twocolumngrid

\setcounter{equation}{0}
\setcounter{figure}{0}

Due to YbMgGaO$_4$ \cite{sMM,sChen1,sChen2,sChen3}, 
easy-plane $XXZ$  $J_1-J_2$ antiferromagnetic model on a triangular lattice 
with additional anisotropic pseudo-dipolar spin-spin interactions 
is of recent interest. We discuss the phase diagram of this model and some aspects of its dynamical response
using SWT approximation and DMRG.

\vspace{-0.5cm}
\subsection{Intuitive derivation of the Hamiltonian}
\vskip -0.3cm

\begin{figure}[t]
\includegraphics[width=0.7\linewidth]{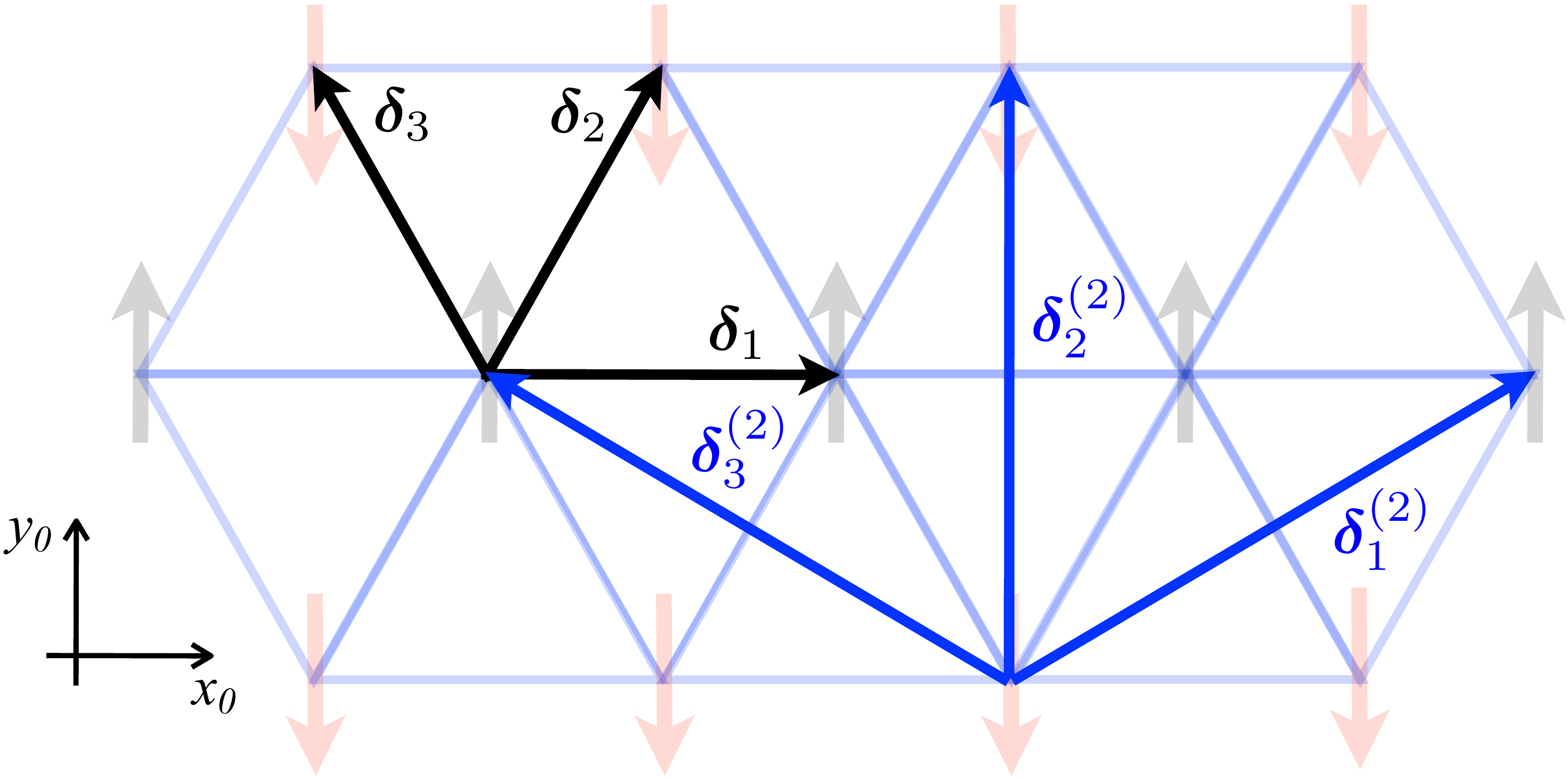}
\vskip -0.2cm
\caption{Nearest and next-nearest primitive vectors.}
\label{s_Fig1}
\vskip -0.5cm
\end{figure}

Consider the most general form of the two-site spin-spin interaction on the ${\bm \delta}_1$ bond in Fig.~\ref{s_Fig1}
with $x_0\parallel {\bm \delta}_1$
\begin{eqnarray}
\hat{\cal H}_{12}={\bf S}^0_1 \left( \begin{array}{ccc} 
J_{xx} & J_{xy} & J_{xz} \\ 
J_{yx} & J_{yy} & J_{yz} \\
J_{zx} & J_{zy} & J_{zz} 
\end{array}\right) {\bf S}^0_2,
\label{H12}
\end{eqnarray}
\vskip -0.15cm \noindent
where ${\bf S}^0=\left(S^{x_0},S^{y_0},S^{z_0}\right)$. The 180$^{\degree}$ rotation around the ${\bm \delta}_1$ bond 
changes $y_0\rightarrow -y_0$, $z_0\rightarrow -z_0$, but should leave the two-site form (\ref{H12}) invariant, leaving
us with
\begin{eqnarray}
\hat{\cal H}_{12}={\bf S}^0_1 \left( \begin{array}{ccc} 
J_{xx} & 0 & 0 \\ 
0 & J_{yy} & J_{yz} \\
0 & J_{zy} & J_{zz} 
\end{array}\right) {\bf S}^0_2.
\label{H12a}
\end{eqnarray}
\vskip -0.15cm \noindent
Inversion with respect to the bond center and changing 
$1\!\leftrightarrow\! 2$ should also  leave (\ref{H12a}) invariant, allowing only symmetric off-diagonal term, 
$J_{zy}\!=\!J_{yz}$. Renaming it $J_{zy}\!=\!J_{z\pm}$, and rewriting the diagonal terms using 
$J_{zz}\!=\!\Delta\cdot J_1$, 
\begin{eqnarray}
J_1=(J_{xx}+J_{yy})/2, \ \ J_{\pm\pm}=(J_{xx}-J_{yy})/4,
\label{Jparam}
\end{eqnarray}
\vskip -0.15cm \noindent
yields the two-site Hamiltonian for ${\bm \delta}_1$
\begin{eqnarray}
&&\hat{\cal H}_{12}=J_1\Big(\Delta S^{z_0}_1 S^{z_0}_2+S^{x_0}_1 S^{x_0}_2+S^{y_0}_1 S^{y_0}_2\Big)
\label{eqH_1b}\\
&&+2J_{\pm\pm}\Big(S^{x_0}_1 S^{x_0}_2-S^{y_0}_1 S^{y_0}_2\Big)
+J_{z\pm}\Big(S^{z_0}_1 S^{y_0}_2+S^{y_0}_1 S^{z_0}_2\Big),
\nonumber
\end{eqnarray}
\vskip -0.15cm \noindent
which clearly follows the structure in  Refs.~\cite{sChen1,sChen2}. The first term is in the familiar $XXZ$ form
and the other two are referred to as pseudo-dipolar terms as they favor spin directions to be (anti)pinned 
to the bond direction \cite{sMZh12}. 

For the other bonds,  and for simplicity, we drop the $J_{z\pm}$ term as it will be ignored later
and consider only the in-plane spin components $S^{x_0}$ and $S^{y_0}$. 
Using the invariance to a $\pi/3$ rotation around the $z_0$ axis 
to transform (\ref{H12a}) to the ${\bm \delta}_2$ bond in Fig.~\ref{s_Fig1}, 
changes $\hat{\bm J}_1$ matrix in  (\ref{H12a}) to
\begin{equation}
\hat{\bm J}_2=\mathbf{R}^{-1}_{\pi/3} \hat{\bm J}_1 \mathbf{R}_{\pi/3}, \ \ 
\mathbf{R}_\theta=\left( \begin{array}{cc} 
\cos\theta & \sin\theta \\ 
-\sin\theta & \cos\theta
\end{array}\right),
\end{equation}
\vskip -0.15cm \noindent
or, explicitly, for the two-component spins using (\ref{Jparam})
\begin{eqnarray}
\hat{\bm J}_2 = \left( \begin{array}{cc} 
J_1+ 2J_{\pm\pm}\cos \frac{2\pi}{3}& 2J_{\pm\pm}\sin \frac{2\pi}{3}\\ 
2J_{\pm\pm}\sin \frac{2\pi}{3} & J_1- 2J_{\pm\pm}\cos \frac{2\pi}{3}
\end{array}\right) .
\label{eqH_2}
\end{eqnarray}
\vskip -0.15cm \noindent
For the bond ${\bm \delta}_3$  in Fig.~\ref{s_Fig1}, rotation is by $2\pi/3$  and 
\begin{eqnarray}
\hat{\bm J}_3 = \left( \begin{array}{cc} 
J_1+ 2J_{\pm\pm}\cos \frac{4\pi}{3}& 2J_{\pm\pm}\sin \frac{4\pi}{3}\\ 
2J_{\pm\pm}\sin \frac{4\pi}{3} & J_1- 2J_{\pm\pm}\cos \frac{4\pi}{3}
\end{array}\right) .
\label{eqH_3}
\end{eqnarray}
\vskip -0.15cm \noindent
Using the auxiliary phases associated with the bond direction ${\bm \delta}_\alpha$  according to 
$\tilde{\varphi}_\alpha\!=\!\{0,-2\pi/3,2\pi/3\}$ as  in Refs.~\cite{sChen1,sChen2}, 
the two-site Hamiltonians in (\ref{eqH_1b}), (\ref{eqH_2}), and (\ref{eqH_3}) can be all reconciled as 
(minus the $J_{z\pm}$ term) 
\begin{eqnarray}
\hat{\cal H}&=&J_1\sum_{\langle ij\rangle}\Big(\Delta S^{z_0}_i S^{z_0}_j+S^{x_0}_i S^{x_0}_j+S^{y_0}_i S^{y_0}_j\Big)
\nonumber\\
&+&2J_{\pm\pm}\sum_{\langle ij\rangle}\Big(S^{x_0}_i S^{x_0}_j-S^{y_0}_i S^{y_0}_j\Big)
\cos\tilde{\varphi}_\alpha
\label{eqH_all}\\
&&\quad\quad \ \ -\Big(S^{x_0}_i S^{y_0}_j+S^{y_0}_i S^{x_0}_j\Big)\sin\tilde{\varphi}_\alpha \, .
\nonumber
\end{eqnarray}
\vskip -0.15cm \noindent
Having in mind effects of disorder discussed below, we note that the 
variations of the diagonal elements of the exchange matrix (\ref{H12a}) by $\pm\delta J_1$, 
translate into variations of $J_{\pm\pm}$ by $\pm \delta J_1/2$ (\ref{Jparam}), which can exceed its bare value. 

\vspace{-0.3cm}
\subsection{$J_1-J_2$ $XXZ$ model}
\vspace{-0.2cm}
Given the hierarchy of terms in YbMgGaO$_4$ \cite{sMM} ($J_{\pm\pm}\!\ll\!J_1$), 
we first neglect  the pseudo-dipolar terms and consider the $J_1-J_2$ $XXZ$ model
\begin{eqnarray}
\hat{\cal H}&=&\sum_{\langle ij\rangle_n}J_n\Big(\Delta S^{z_0}_i S^{z_0}_j+S^{x_0}_i S^{x_0}_j+S^{y_0}_i S^{y_0}_j\Big)
\label{eqHJ1J2}
\end{eqnarray}
\vskip -0.15cm \noindent
where  the sums are over the nearest- and next-nearest-neighbor   
bonds, $J_{1(2)}\!>\!0$, same anisotropy $0\!\leq\Delta\!\leq\!1$ is assumed for both couplings, 
and $\{x_0,y_0,z_0\}$ is the laboratory reference frame, see Fig.~\ref{s_Fig1}.
Triangular-lattice primitive vectors (in units of lattice spacing $a$) are
$\bm{\delta}_1 = (1,0)$, and $\bm{\delta}_{2(3)} = (\pm1/2, \sqrt{3}/2)$.
The next-nearest neighbor sites also form triangular lattices  with the vectors 
$\bm{\delta}^{(2)}_{1(3)} = (\pm3/2,\sqrt{3}/2)$ and $\bm{\delta}^{(2)}_2 = (0, \sqrt{3})$.

\begin{figure}[b]
\vskip -0.3cm
\includegraphics[width=\linewidth]{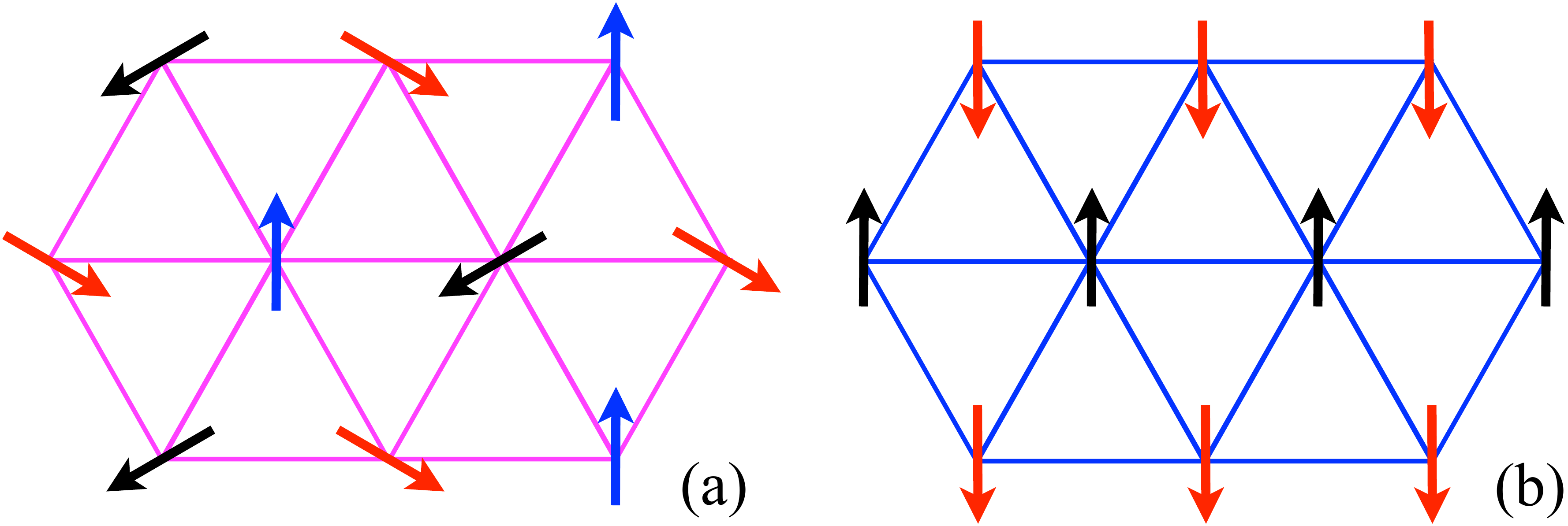}
\vskip -0.2cm
\caption{(a) The  $120^{\degree}$, and (b) the collinear (stripe) states. }
\label{s_Fig2}
\end{figure}

Because of the $XXZ$ anisotropy, the spins are in the $x_0$-$y_0$ plane.
Performing a standard rotation to the local reference frames of individual spins 
and confining ourselves to the leading  $1/S$-order terms yields classical energy and harmonic SWT Hamiltonian.
The two states compete for $0\!<\!\alpha\!=\!J_2/J_1\!<\!1$, the $120^{\degree}$
and the collinear (``stripe'') states, where in the latter rows of ferromagnetically ordered spins align antiferromagnetically, 
see Fig.~\ref{s_Fig2}. While the situation is somewhat more subtle 
for the selection of the stripe phase, see \cite{sChubukov,sJolicquer}, they suffice for
our consideration. 
For the  $120^{\degree}$ state, $J_2$ couples spins 
on the same sublattices. For the stripe state, choosing the ferromagnetic rows in the $x_0$-direction 
as in Fig.~\ref{s_Fig2}, only two of the $J_2$ bonds are ferromagnetic. 
The classical energies per site of the two states  are 
\vskip -0.15cm \noindent
\begin{eqnarray}
E^{120^{\degree}}_{\rm cl}/S^2=-3J_1/2+3J_2, \ \ 
E^{\rm coll}_{\rm cl}/S^2=-J_1-J_2, 
\label{eqEcl}
\end{eqnarray}
\vskip -0.15cm \noindent
yielding a transition between the two at $\alpha_c=1/8$, independently of $\Delta$.

\vspace{-0.4cm}
\subsubsection{Linear spin-wave theory}
\vspace{-0.2cm}

In both phases, one obtains the harmonic Hamiltonian in a standard form
\vskip -0.15cm \noindent
\begin{equation}
\hat{\cal H}^{(2)}=3J_1S\sum_{{\bf k}} 
A_{\bf k}  a^\dagger_{{\bf k}}a^{\phantom{\dag}}_{{\bf k}}   
-\frac{B_{\bf k}}{2} \left( a^\dagger_{{\bf k}} a^{\dagger}_{-{\bf k}}
+{\rm H.c.}\right), \label{eq_H2}
\end{equation}
\vskip -0.15cm \noindent
with the  Bogolyubov transformation giving magnon energy  $\varepsilon_{{\bf k}}\!=\!3J_1S \omega_{{\bf k}}$ with
$\omega_{{\bf k}}=\sqrt{A_{\bf k}^2-B_{\bf k}^2}$, 
and the ordered magnetic moment
\vskip -0.15cm \noindent
\begin{eqnarray}
\langle S\rangle=S-\frac12\sum_{\bf k} \left(\frac{A_{\bf k}}{\omega_{\bf k}}-1\right).
\label{eq_dS}
\end{eqnarray}
\vskip -0.15cm \noindent
Parameters for the $120^{\degree}$ phase are
\vskip -0.15cm \noindent
\begin{eqnarray}
&&A_{\bf k}=1+\left( \Delta-1/2\right)\gamma_{{\bf k}}-
\alpha\left(2-\left( \Delta+1\right)\gamma_{{\bf k}}^{(2)}\right) , \quad\\
&&B_{\bf k}=-\gamma_{{\bf k}}\left( \Delta+1/2\right) -
\alpha\left( \Delta-1\right)\gamma_{{\bf k}}^{(2)},
\label{eq_ABk120}
\end{eqnarray}
\vskip -0.15cm \noindent
and for the stripe phase
\vskip -0.15cm \noindent
\begin{eqnarray}
&&\bar{A}_{\bf k}=2/3+\Delta\gamma_{{\bf k}}+\gamma_{{\bf k}}^{\prime}+
\alpha\left(2/3+\Delta\gamma_{{\bf k}}^{(2)}+\gamma_{{\bf k}}^{\prime(2)}\right) , \quad\\
&&\bar{B}_{\bf k}=\gamma_{{\bf k}}^{\prime}-\Delta\gamma_{{\bf k}}+
\alpha\left( \gamma_{{\bf k}}^{\prime(2)}-\Delta\gamma_{{\bf k}}^{(2)}\right),
\label{eq_ABk_coll}
\end{eqnarray}
\vskip -0.15cm \noindent
where we used the amplitudes 
\vskip -0.15cm \noindent
\begin{eqnarray}
&&\gamma_{{\bf k}} \left[\gamma_{{\bf k}}^\prime \right]
=\frac{1}{3}\Big(\cos k_x\pm 2\cos\frac{k_x}{2}\cos \frac{\sqrt{3}k_y}{2}\Big),\\
&&\gamma_{{\bf k}}^{(2)}\left[\gamma_{{\bf k}}^{\prime(2)}\right] 
=\frac{1}{3} \Big(\cos \sqrt{3}k_y\pm 2\cos\frac{3k_x}{2}\cos \frac{\sqrt{3}k_y}{2}\Big).\nonumber
\end{eqnarray}
\vskip -0.15cm \noindent

\begin{figure}[t]
\includegraphics[width=0.8\linewidth]{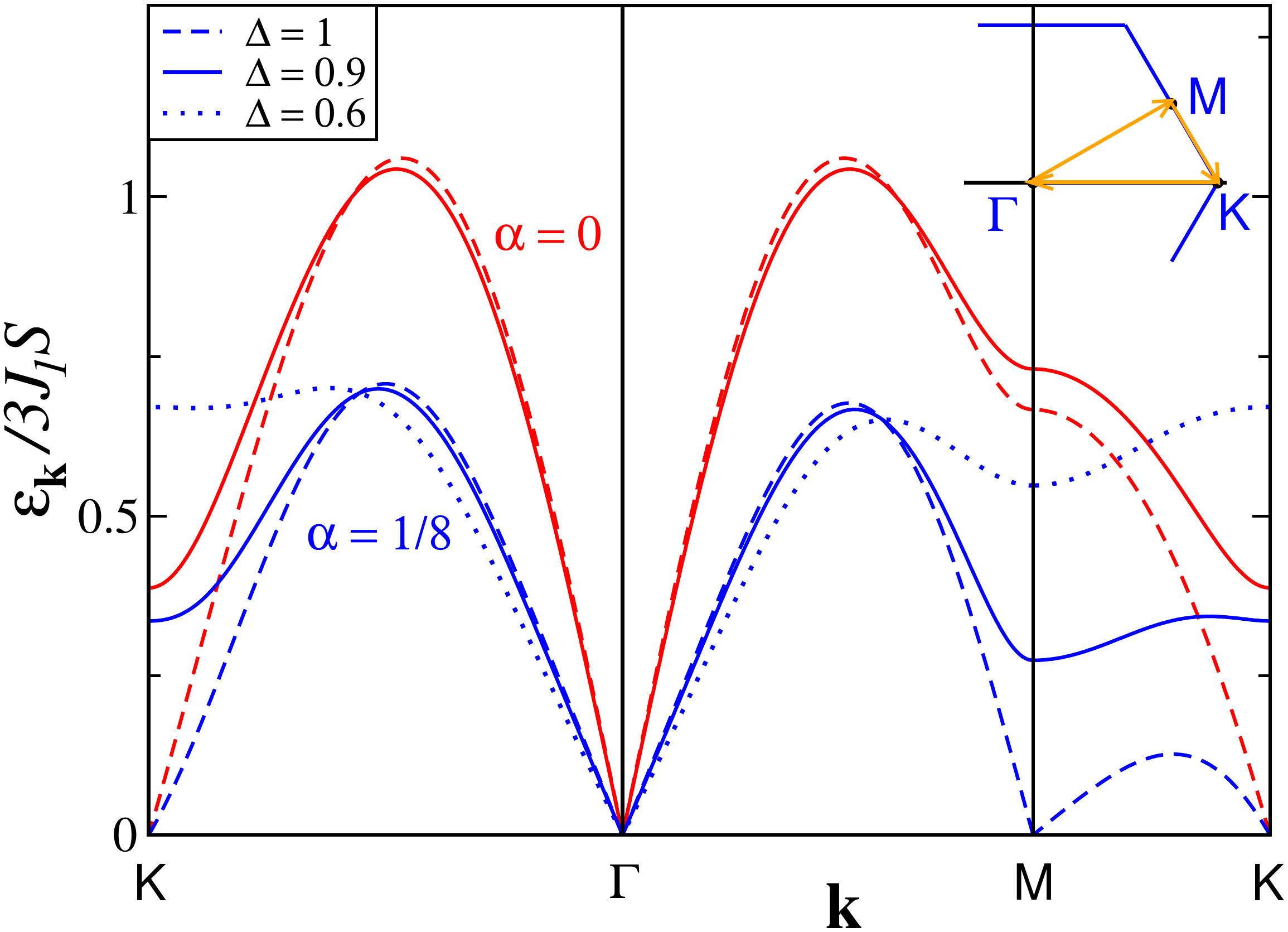}
\vskip -0.2cm
\caption{$\omega_{\bf k}$ in the  $120^{\degree}$ phase for several  $\alpha$ and $\Delta$.}
\label{s_wk120}
\vskip -0.5cm
\end{figure}

\vspace{-0.5cm}
\subsubsection{Magnon spectra}
\vspace{-0.2cm}

Since $\Delta\!<\!1$ leaves only $U(1)$ symmetry intact, the spectra in both phases should exhibit only one 
Goldstone mode. This is the case for the  $120^{\degree}$ phase, see Fig.~\ref{s_wk120}. At the 
Heisenberg limit ($\Delta\!=\!1$), the Goldstone modes are at both $\Gamma$ and K(K$^\prime$) points 
[dashed lines]
with an additional zero-mode occurring at the M points at the transition to the stripe phase,  $\alpha\!=\!1/8$.   
Since gaps grow  with $1\!-\!\Delta$, one can expect lower  quantum fluctuations  in the ground state and 
a more robust magnetic order for $\Delta\!<\!1$. 

\begin{figure}[t]
\includegraphics[width=0.8\linewidth]{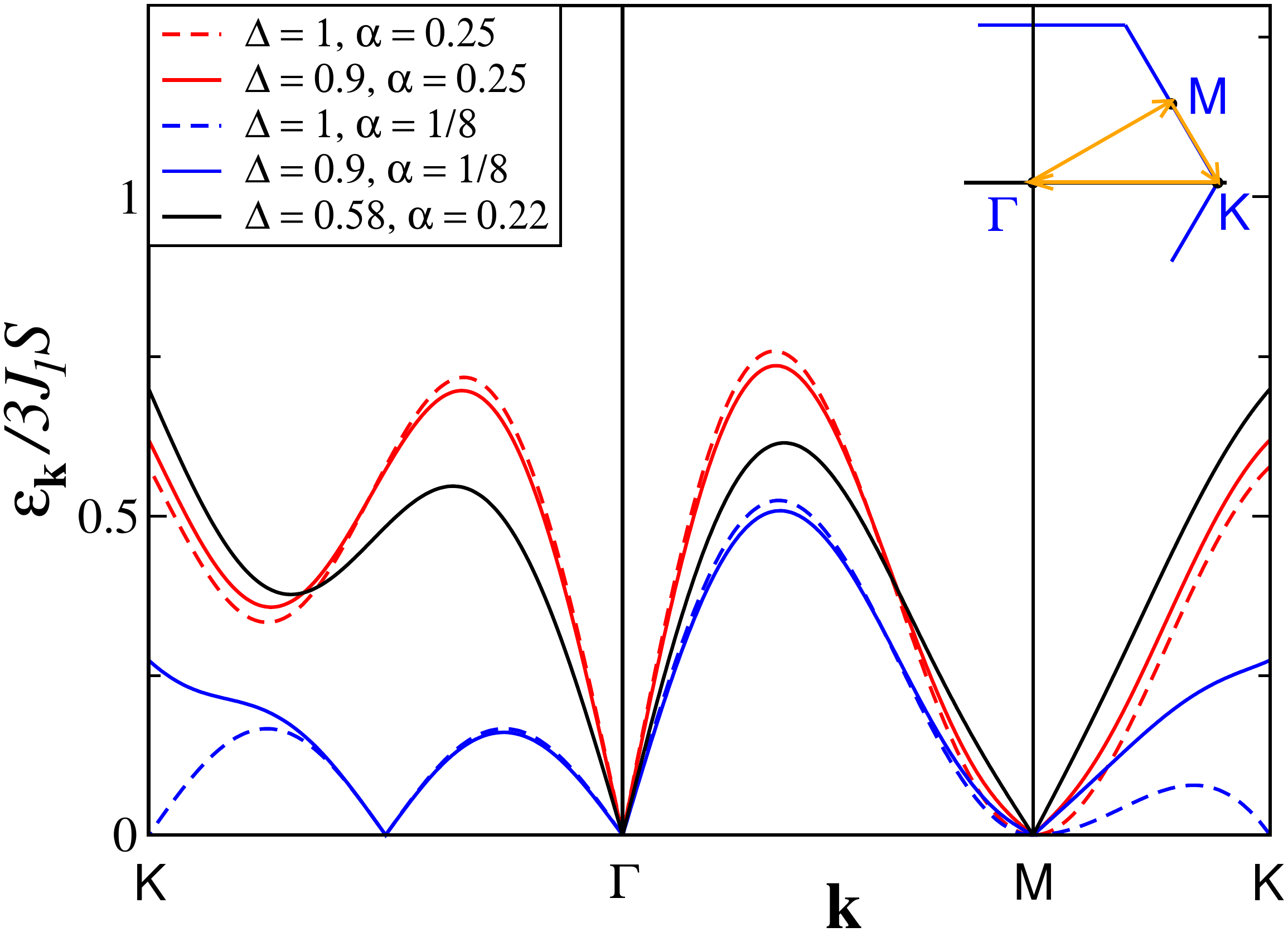}\\
\includegraphics[width=0.8\linewidth]{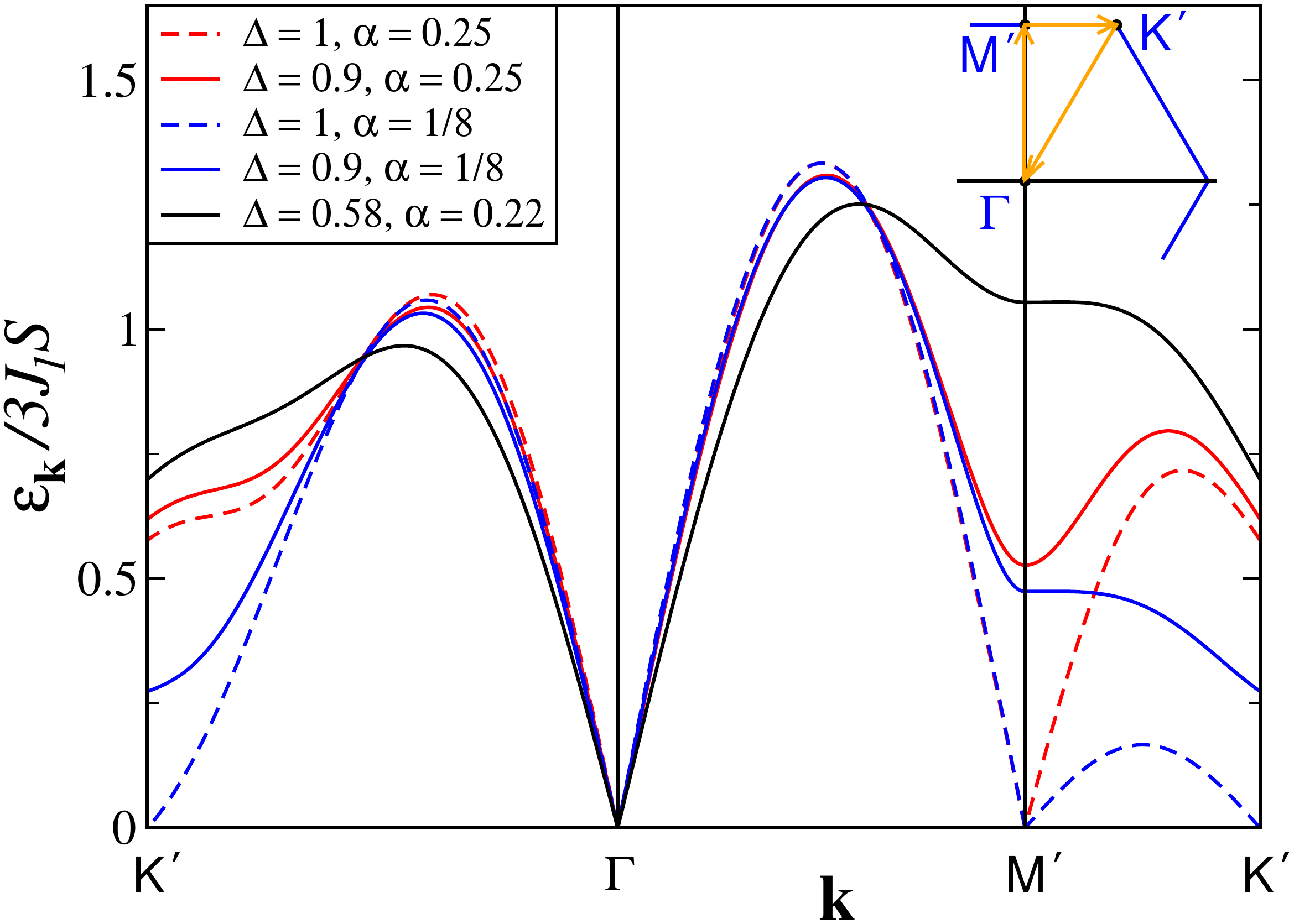}
\vskip -0.3cm
\caption{$\omega_{\bf k}$ in the  stripe phase for several  $\alpha$ and $\Delta$.}
\label{s_wk_coll}
\vskip -0.6cm
\end{figure}

The situation is more involved for the stripe phase  where extra zero modes occur due
to accidental degeneracy. 
The latter should be lifted by the fluctuations in the next order, see Refs.~\cite{sChubukov,sJolicquer}. 
In Fig.~\ref{s_wk_coll}, we present $\omega_{{\bf k}}$ for the stripe state shown in Fig.~\ref{s_Fig1}, for 
which the ordering vector is M$^\prime$ with the mode at M points being accidental. 
Since the symmetry of the stripe state is different, 
the $\Gamma$KM$\Gamma$ and $\Gamma$K$^\prime$M$^\prime\Gamma$
directions are not equivalent.

One ``dangerous'' feature of the accidental zero modes is that the magnon dispersion near them for $\Delta\!=\!1$
and any $\alpha>1/8$ is $\omega_{{\bf k}}\propto k^2$. One can expect this to lead to logarithmically 
divergent fluctuation corrections \cite{sChubukov}, e.g., to the order parameter (\ref{eq_dS}). However, we find 
that the fluctuation parts of the Bogolyubov transformation, $v_{\bf k}$, vanish along the $(k_x,\pm {\rm M}^\prime/2)$ 
lines that contain accidental zero modes, thus avoiding the divergence. 
Such vanishing of fluctuations is related to the ferromagnetic-like ordering along the $x$-direction.
This feature allows us to stay within the linear SWT for the stripe state. 

$XXZ$ anisotropy gaps out zero modes at K, K$^\prime$, and M$^\prime$ 
and modifies the dispersion of the accidental modes to 
a more standard $\omega_{{\bf k}}\propto k$. 
Overall, the trend is the same: $\Delta<1$ reduces the role 
of quantum fluctuations and makes magnetic order more stable. In Fig.~\ref{s_wk_coll}, 
we also plot the dispersion for $J_2/J_1\!=\!0.22$ and $\Delta\!=\!0.58$ that are experimentally relevant to YbMgGaO$_4$, 
\cite{sMM}.

\vspace{-0.5cm}
\subsubsection{$XXZ$ phase diagram}
\vspace{-0.3cm}

From the character of the magnon spectra on both sides of the $\alpha=1/8$ transition, one can see 
one important difference of the $J_1-J_2$ problem on the triangular lattice from the square-lattice and other 
 counterparts. Here, this transition is not associated with a divergence in the spin-wave fluctuation corrections.
For instance, $\delta S$ in (\ref{eq_dS}) remains finite. This implies that at $S\gg 1$ the transition between
the 120$^{\degree}$ and stripe phase must become a direct one for any $\Delta$. In fact, within the SWT,
the magnetically disordered state disappears already for $S\!=\!1$, in agreement with \cite{sIvanov}.

\begin{figure}[t]
\includegraphics[width=\linewidth]{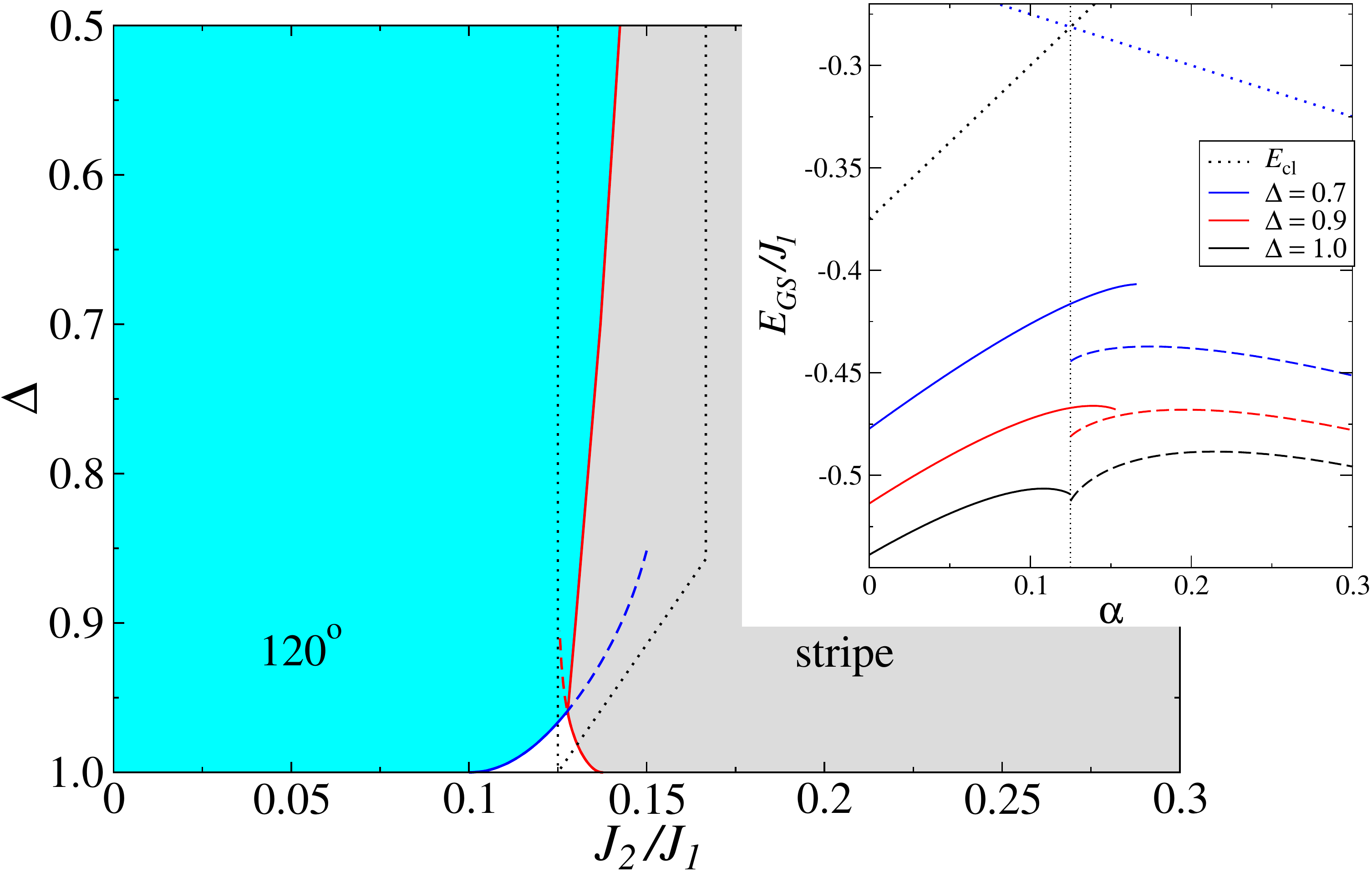}
\vskip -0.2cm
\caption{Phase diagram $\Delta$ vs $\alpha$. Right: $E_{GS}$ vs $\alpha$.}
\label{s_PhD1}
\vskip -0.5cm
\end{figure}

Another important feature of the problem, is that  for $\Delta\!<\!1$ the spectrum of the 120$^{\degree}$ state
remains well-defined beyond the classical $\alpha_c\!=\!1/8$  \cite{sIvanov}. For $\Delta\!<\!6/7$, 
it extends to $\tilde{\alpha}_c\!=\!1/6$, and for $6/7\!<\!\Delta\!<\!1$ it is a straight line
connecting $\alpha_c$ and $\tilde{\alpha}_c$, see Fig.~\ref{s_PhD1}. Thus, the stability regions of the spectra 
of the two phases always overlap for $\Delta\!<\!1$, thus favoring a direct transition between them. 

\begin{figure}[b]
\vskip -0.5cm
\includegraphics[width=0.9\linewidth]{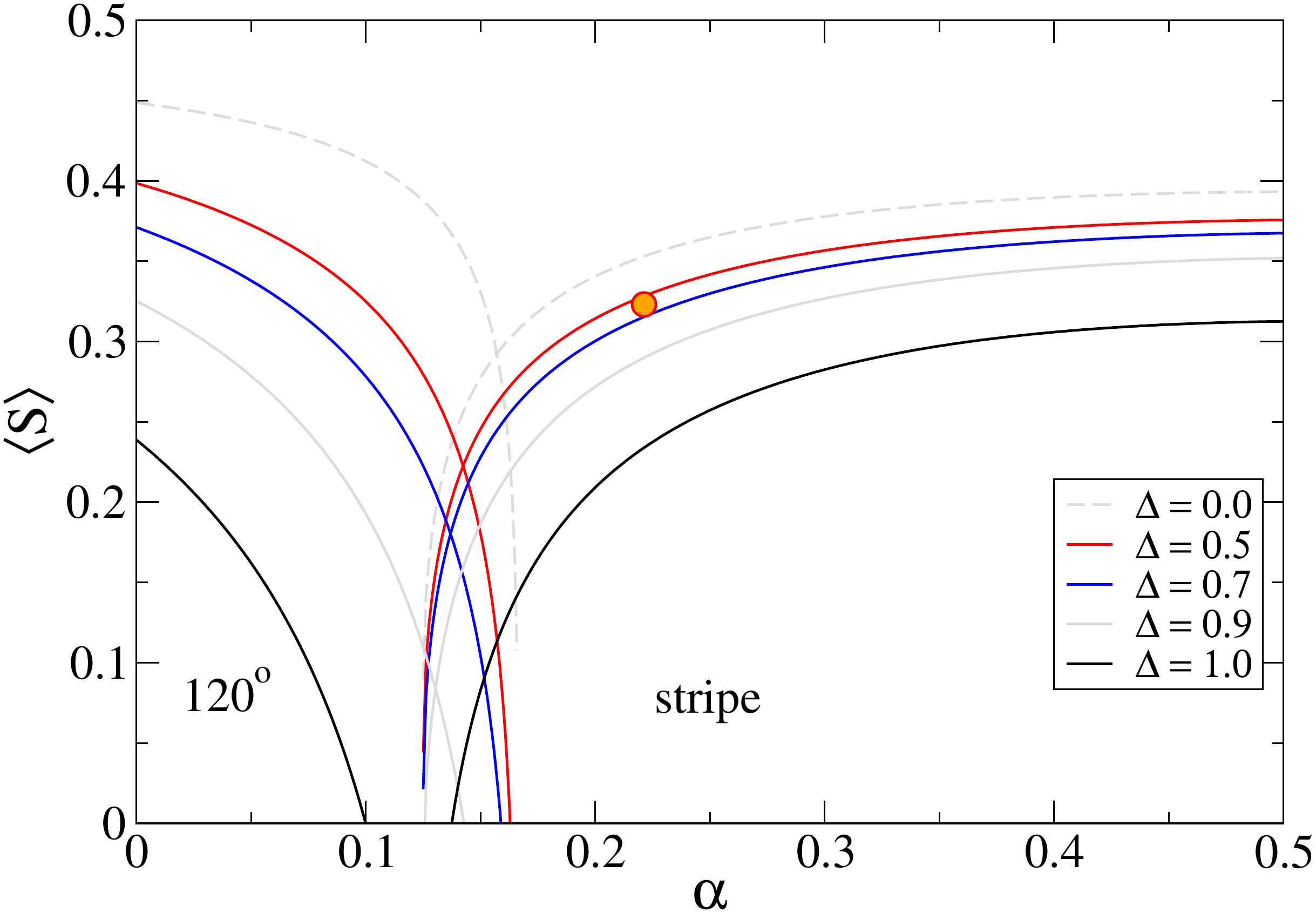}
\vskip -0.2cm
\caption{$\langle S\rangle$ vs $\alpha$.}
\label{s_dS_vs_alpha_Delta}
\end{figure}

In Fig.~\ref{s_PhD1} we present the SWT $\alpha-\Delta$ phase diagram of the $XXZ$ model (\ref{eqHJ1J2}) 
for $S\!=\!1/2$. It is obtained  from the $\alpha$- and $\Delta$-dependencies
of the ordered moment (\ref{eq_dS}), with their representatives 
shown in Fig.~\ref{s_dS_vs_alpha_Delta}. The boundary of each phase is found from the
$\langle S\rangle\!=\!0$ condition. The direct transition between them is inferred from their magnetization crossings
to signify the region where there is no intermediate non-magnetic state. The determination of such a transition 
from the crossings of the groundstate (GS) energies of the competing states 
meets rather standard difficulty within the SWT.
Quantum corrections split the GS energies of the two phases, with no opportunity for a crossing in the region where 
both spectra are well-defined, see Fig.~\ref{s_PhD1}. However, one can still infer from Fig.~\ref{s_PhD1} that
the  energy  shift is more significant for the stripe phase, 
indicating that the transition  is likely to shift toward  smaller values of $\alpha$ for smaller $\Delta$.

Thus, the $XXZ$ anisotropy favors ordered states  and, for the 
YbMgGaO$_4$ parameters ($J_2/J_1\!=\!0.22$ and $\Delta\!=\!0.58$,
see \cite{sMM}), SWT places it in a stripe phase with a large ordered moment
$\langle S\rangle\!\approx \! 0.32$, see orange dot in Fig.~\ref{s_dS_vs_alpha_Delta}.

\vspace{-0.5cm}
\subsection{Pseudo-dipolar terms}
\vspace{-0.3cm}

The pseudo-dipolar terms are introduced in (\ref{eqH_all}) and 
we omit the couplings of $S^{x_0(y_0)}$ to the out-of-plane spin components $S^{z_0}$, referred to as 
the $J_{z\pm}$ terms, as those are negligible in YbMgGaO$_4$, see \cite{sMM,sChen1}.

For the spins in the coplanar configuration with the single ordering vector ${\bf Q}$,  one can 
transform (\ref{eqH_all}) to the local reference frames  with the $z$-axes  along the spin quantization axes and obtain 
the contribution of the pseudo-dipolar  terms to the classical energy 
\begin{eqnarray}
\delta E_{\rm cl}\!=\! J_{\pm\pm} S^2 \sum_{i,\pm\alpha}\cos\left(2\varphi_0\! +\! 2{\bf Q}\cdot{\bf r}_i
\!+\!{\bf Q}\cdot{\bm\delta}_{\pm\alpha}\!+\!\tilde{\varphi}_\alpha
\right). \ \ 
\label{eqEJpm}
\end{eqnarray}
\vskip -0.15cm \noindent
where $\tilde{\varphi}_\alpha$ is the auxiliary bond-dependent phase factor for the ${\bm \delta}_\alpha$ bonds, 
see Fig.~\ref{s_Fig1},  with  $\tilde{\varphi}_\alpha\!=\!\{0,-2\pi/3,2\pi/3\}$,
$i\pm\alpha={\bf r}_i\pm{\bm\delta}_\alpha$,  
and $\varphi_0$ 
is the spin direction relative to the $x_0$ axis  at a reference site $i=0$. 

\emph{Stripe phase.---}%
One can check whether the pseudo-dipolar terms favor deviations of the spins away from the 
stripe order. This is done by considering $\hat{\cal H}_{\rm p-d}$ in 
(\ref{eqH_all}) and showing that the terms linear in the spin deviations vanish.
Indeed, for any of the stripe states, on one of the bonds such linear terms vanish and the 
``tug'' on the two other bonds cancels out. Thus, the the pseudo-dipolar terms leave the stripe state stable.

\begin{figure}[b]
\vskip -0.5cm
\includegraphics[width=0.8\linewidth]{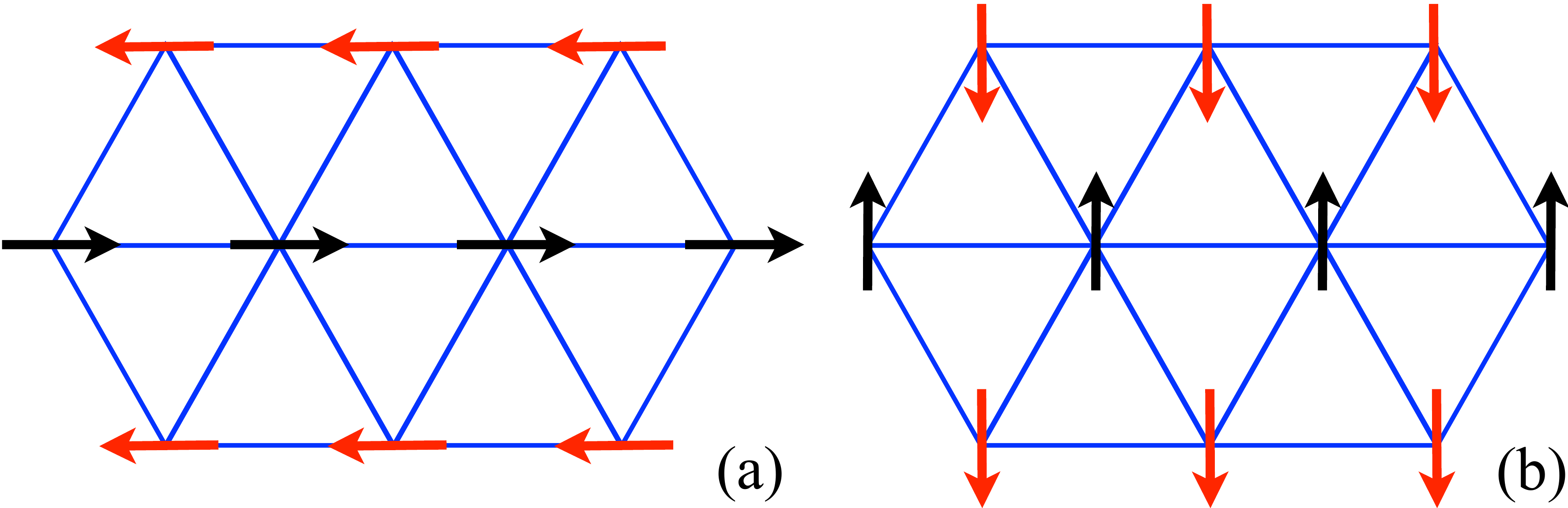}
\vskip -0.2cm
\caption{Spin orientation in the stripe phase with ${\bf Q}=M'$ and pseudo-dipolar terms 
with (a) $J_{\pm\pm}<0$ and (b) $J_{\pm\pm}>0$.}
\label{s_Jpm_pin}
\end{figure}

One can see that $2{\bf Q}\cdot{\bf r}_i\!=\!2\pi n$ for ${\bf Q}$ at any of the $M$ points (stripe ordering vectors).
The  ${\bf Q}\cdot{\bm\delta}_{\pm\alpha}$ phase factors are either 0 or $\pm\pi$ for the three 
primitive vectors ${\bm\delta}_{\alpha}$ with their values dependent on the choice of ${\bf Q}$. That is, 
${\bf Q}\cdot{\bm\delta}_{\alpha}\!=\!\{\pi,\pi,0\}$ for  ${\bf Q}\!=\!M$, $\{0,\pi,\pi\}$ for  ${\bf Q}\!=\!M'$, 
and $\{-\pi,0,\pi\}$ for  ${\bf Q}\!=\!M''$.
Thus the classical energy contribution of the pseudo-dipolar  terms simplifies to 
\begin{eqnarray}
\delta E_{\rm cl}=2J_{\pm\pm} S^2N \sum_{\alpha}\cos\left(2\varphi_0 
+{\bf Q}\cdot{\bm\delta}_{\alpha}+\tilde{\varphi}_\alpha
\right). \ 
\label{eqEJpm1}
\end{eqnarray}
\vskip -0.15cm \noindent
Minimization of it with respect to the ``global'' spin angle $\varphi_0$ is expected to ``pin'' the orientation
of the stripe spin structure in Fig.~\ref{s_Fig2}(b) to the lattice.  
Using the auxiliary phase factors 
$\tilde{\varphi}_\alpha$, one 
can find that the energy minimum is reached when the bonds  ${\bm\delta}_{\alpha}\perp{\bf Q}$ 
are satisfied completely ($\cos\theta_{\alpha} \!=\!-1$) while the other two bonds are partially satisfied 
($\cos\theta_{\alpha} \!=\!-1/2$). This translates 
to the energy contribution $\delta E_{\rm cl}\!=\!-4|J_{\pm\pm}| S^2N$ and the spins' global orientation 
either parallel ($J_{\pm\pm}\!<\!0$) or perpendicular ($J_{\pm\pm}\!>\!0$) to the ``happy'' bond, depending on the 
sign of $J_{\pm\pm}$. To be specific, for the choice of ${\bf Q}\!=\!M'$  as in Fig.~\ref{s_Fig2}, 
the pseudo-dipolar terms will pin the spin orientation  along the $x_0$-axis (${\bm\delta}_1$ bond) 
for $J_{\pm\pm}\!<\!0$ and along the 
$y_0$-axis if $J_{\pm\pm}\!>\!0$, see Fig.~\ref{s_Jpm_pin}.

\begin{figure}[t]
\includegraphics[width=0.8\linewidth]{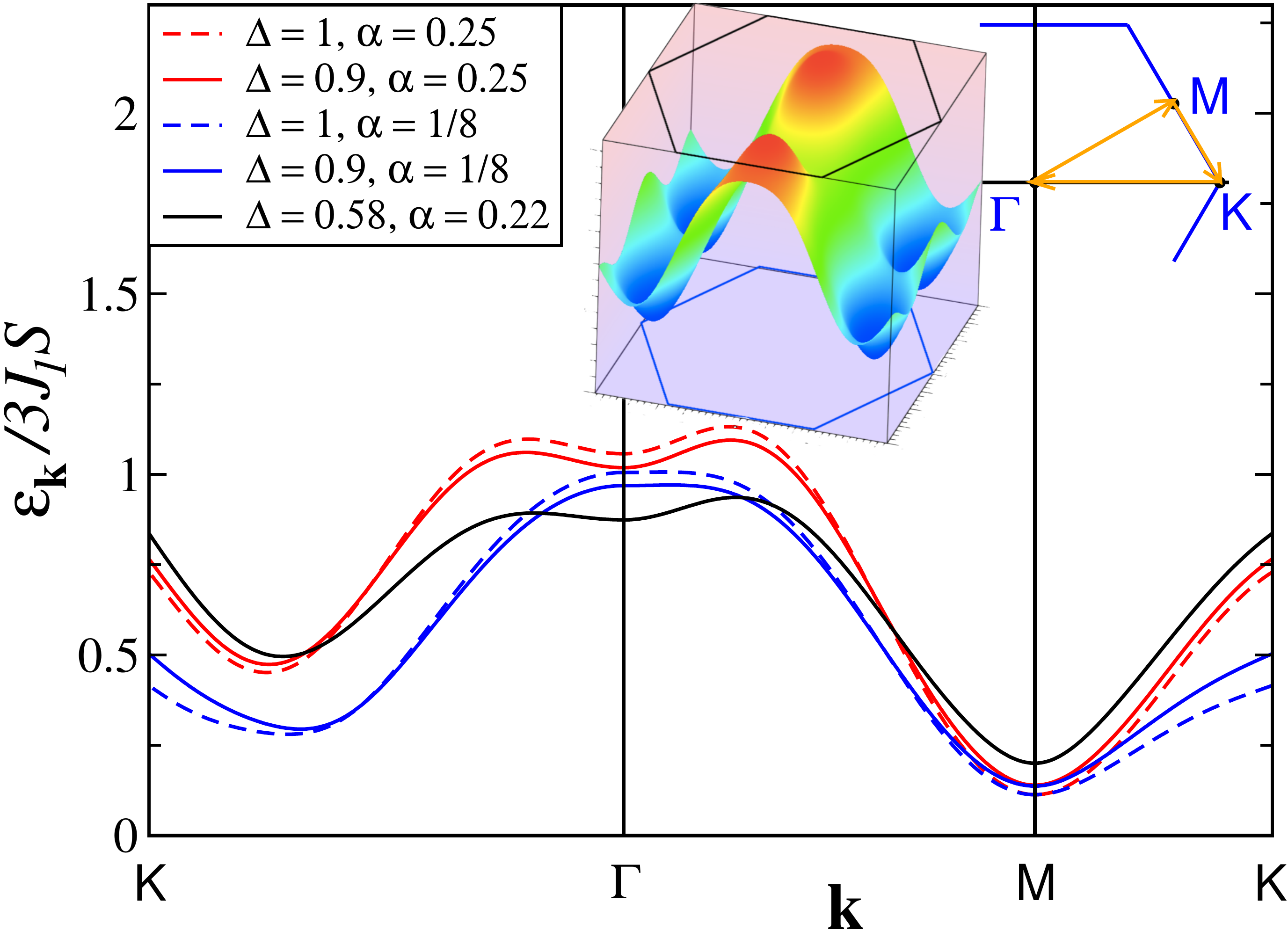}\\
\includegraphics[width=0.8\linewidth]{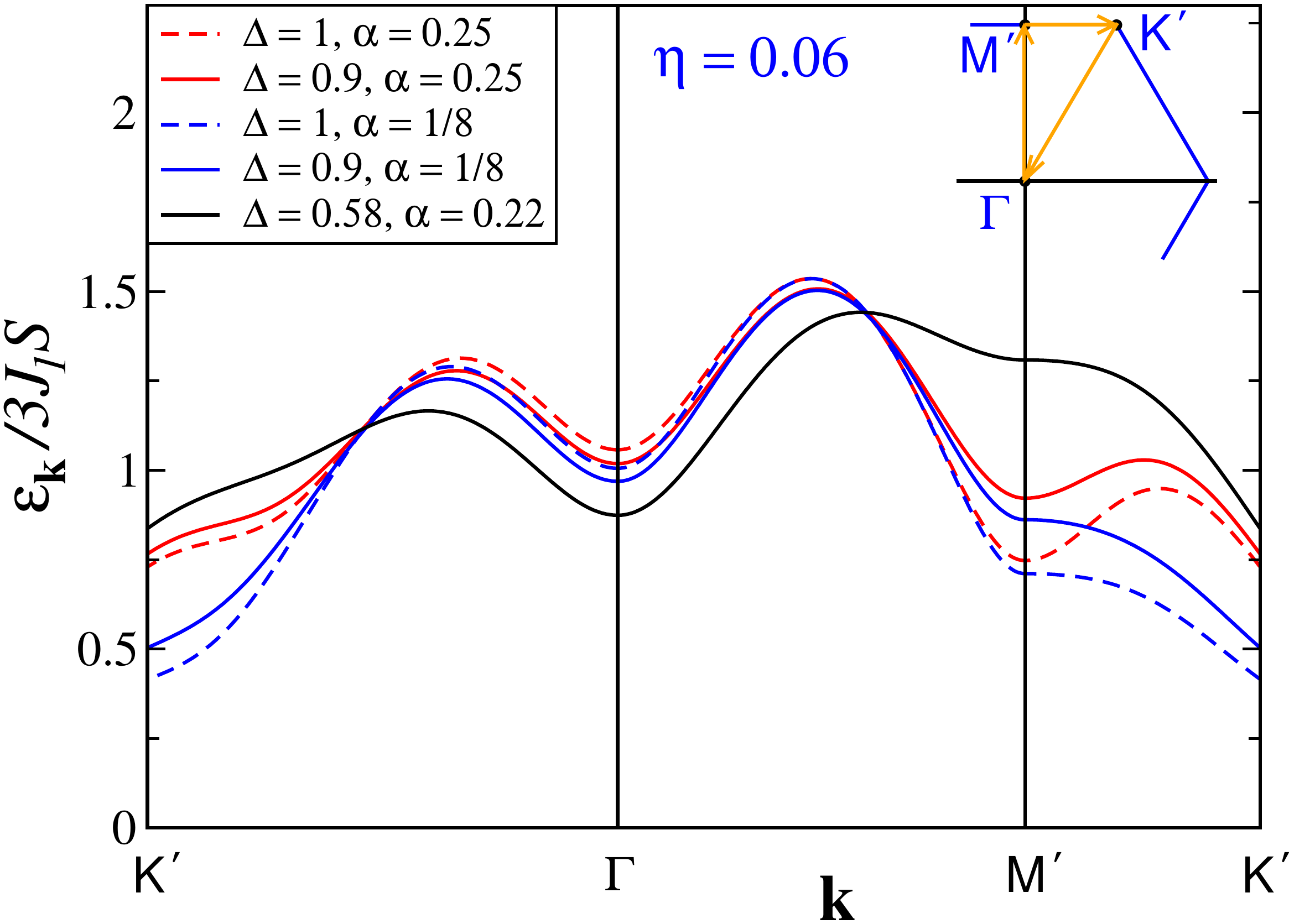}
\vskip -0.2cm
\caption{$\omega_{\bf k}$ in the  stripe phase for the same  $\alpha$ and $\Delta$ as in Fig.~\ref{s_wk_coll}, 
$\eta\!=\!|J_{\pm\pm}|/J_1\!=\!0.06$. Inset and black lines are the same.}
\label{s_wk_collpm}
\vskip -0.5cm
\end{figure}

For the contributions from the pseudo-dipolar terms to the magnon spectrum 
the sign of $J_{\pm\pm}$ does not matter as the  
choice of the  ``global'' spin orientation angle $\varphi_0$ that minimizes energy also 
changes their overall sign to positive. 
Choosing ${\bf Q}=M'$ stripe order in accord with the choice above 
leads to the corrections to the spin-wave parameters in (\ref{eqH_2})
\begin{eqnarray}
\delta\bar{A}_{\bf k}= \frac{8\eta}{3}+\frac{\eta}{2}\left(3\gamma_{{\bf k}}+\gamma_{{\bf k}}^{\prime}\right), \ \
\delta\bar{B}_{\bf k}=\frac{\eta}{2}\left(3\gamma_{{\bf k}}+\gamma_{{\bf k}}^{\prime}\right), \ 
\label{eq_dABk_coll}
\end{eqnarray}
\vskip -0.15cm \noindent
where $\eta\!=\!|J_{\pm\pm}|/J_1$. 
Plots of the magnon energy $\varepsilon_{{\bf k}}$ along the $\Gamma MK\Gamma$ and the $\Gamma M'K'\Gamma$
cuts are shown in Fig.~\ref{s_wk_collpm} for $\eta\!=\!0.06$ and the same sets of parameters $\Delta$ and $\alpha$ as in 
Fig.~\ref{s_wk_coll}. The most significant difference between   Figs.~\ref{s_wk_coll} and \ref{s_wk_collpm} is the 
opening of sizable gaps and
lifting of the accidental degeneracy mode at the $M$ point. These effects are due to a low  
symmetry of the  model with the pseudo-dipolar terms. This should 
strengthen  the stripe order via a reduction of the quantum fluctuations. 
 
\emph{$120^{\degree}$ state.---}%
The consideration of the effects of the pseudo-dipolar terms  on the $120^{\degree}$ state is 
somewhat more involved. One can show that contributions of the three 
sublattices to the classical energy cancel each other and that the terms linear in spin deviations
also vanish. Thus, the $120^{\degree}$ state is locally stable to the pseudo-dipolar terms, 
$\delta E^{120^{\degree}}_{\rm cl}=0$, and 
no pinning of a particular global order parameter orientation to the lattice occurs.

\vspace{-0.5cm}
\subsubsection{Modified phase diagram}
\vspace{-0.3cm}

The most important outcome of the pseudo-dipolar terms is two-fold. First,
the (per site) classical energy of the stripe state is lowered  to 
$E^{\rm coll}_{\rm cl}/S^2\!=\!-J_1-J_2-4|J_{\pm\pm}|$ 
while the energy of the $120^{\degree}$ state is unchanged from (\ref{eqEcl}). 
This leads to an expansion of the stripe phase in the $\Delta\!-\!J_2$ phase diagram with the classical transition 
moved down to $(J_2/J_1)_c\!=\!1/8-\eta$,
completely suppressing the $120^{\degree}$ state at a modest $|J_{\pm\pm}|/J_1\!=\!0.125$. Second, because of the 
lack of continuous symmetries, gapped excitation spectra should reduce quantum fluctuations and diminish
the already suppressed magnetically disordered window of the anisotropic $J_1\!-\!J_2$ $XXZ$ model,
making transition between the $120^{\degree}$ and the stripe phase a direct one for 
the entire range of $\Delta$ at a rather small $J_{\pm\pm}$.

\begin{figure}[t]
\includegraphics[width=0.9\linewidth]{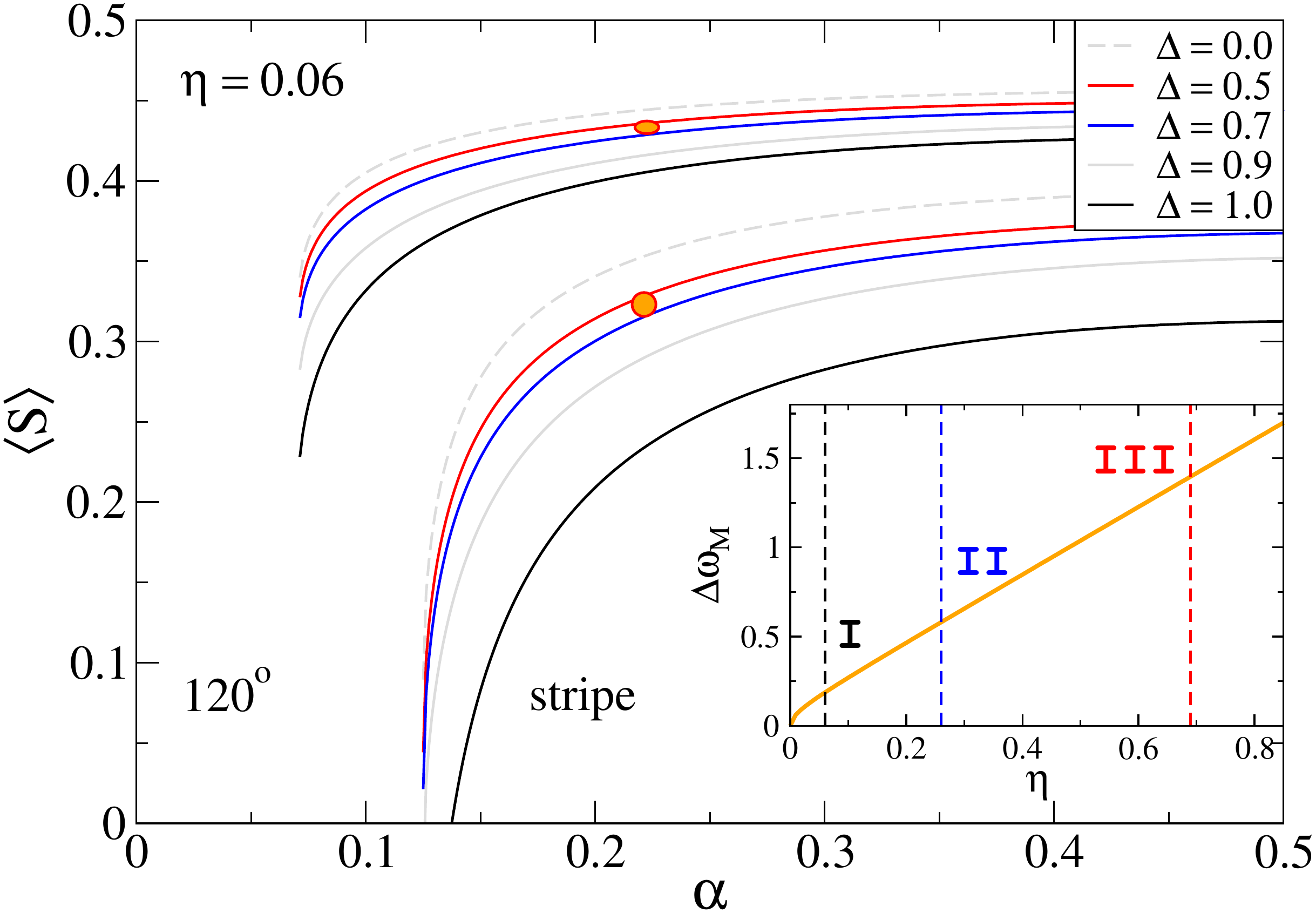}
\vskip -0.2cm
\caption{ $\langle S\rangle$ vs $\alpha=J_2/J_1$ for $\eta=\!|J_{\pm\pm}|/J_1\!=0$ and $\eta=0.06$.
Inset: lowest magnon gap in units of $3J_1S$ vs $\eta$ by SWT.}
\label{s_dS_vs_alpha_Jpm}
\vskip -0.5cm
\end{figure}

We demonstrate these effects in Fig.~\ref{s_dS_vs_alpha_Jpm}, which shows   
the $\alpha$-dependence of the ordered moment for several $\Delta$'s
in the stripe phase. First set is the same as in Fig.~\ref{s_dS_vs_alpha_Delta} for $J_1\!-\!J_2$ $XXZ$ model
with $\eta\!=\!0$ and the second is for $\eta\!=\!0.06$. 
The orange dot marks $J_2/J_1\!=\!0.22$ and $\Delta\!=\!0.58$ as before.
It is clear that the stripe phase 
expands to the lower values of $\alpha$ and the ordered moment is increased, in agreement
with the expectations. The transition in Fig.~\ref{s_dS_vs_alpha_Jpm} is
at the values only slightly larger than the classical value $\tilde{\alpha}_c\!=\!0.065$ via 
an instability of the magnon branch at an incommensurate 
wavevector along the $\Gamma K$ line. 

\emph{Other parameter sets.---}%
There are three sets of   parameters that were inferred from the 
magnon dispersion  of  YbMgGaO$_4$ 
in the high-field phase. First is 
$\Delta\!=\!0.58$, $\alpha\!=\!0.22$, and $\eta\!=\!0.06$  \cite{sMM}, which we 
use in our plots and will refer to as {\tt Set I}. 
Second and third are attempts to fit the same data without 
next-neighbor exchange, so both have $\alpha\!=\!0$ and  
$\Delta\!=\!0.75$ and $\eta\!=\!0.26$ \cite{sChen3} ({\tt Set II}) and  
$\Delta\!=\!0.55$ and $\eta\!=\!0.69$ \cite{sMM} ({\tt Set III}), respectively.

We note that only {\tt Set I} is compatible with the ESR data \cite{sChen1}.
The values of $J_{\pm\pm}$ in {\tt Sets II} and {\tt III} put the system firmly in the stripe phase and
have no $120^{\degree}$ state in a vicinity.
Moreover,  {\tt Sets II} and {\tt III} correspond to much larger gaps in the magnon spectra than  
{\tt Set I}, see inset in Fig.~\ref{s_dS_vs_alpha_Jpm}, which shows the lowest magnon energy 
 vs $\eta$; the SWT expression for the gap is given by 
$\Delta\omega_{\rm M}\!=\!\frac43\sqrt{2\eta(\eta+(1-\Delta)/4)}$. While for YbMgGaO$_4$ {\tt Set I} 
gives the gap  $\sim\!0.06$ meV, which is below experimental energy resolution \cite{sMM}, 
for both  {\tt Set II} and {\tt Set III} the gaps are well above it  and should have been readily observed. 
In addition, the values of the  ordered moment within the SWT
for all three sets are nearly classical:
$\langle S\rangle\!=\!0.433$, $0.456$, and $0.486$, respectively.

 \vspace{-0.5cm}
\subsection{Polarized phase, $H>H_s$, out-of-plane field}
\vspace{-0.3cm}

A strong out-of-plane magnetic field that is sufficient to co-align all the spins, 
should allow, at least in principle, to determine the magnitide of the pseudo-dipolar terms from
several observables \cite{sMM}.

The full Hamiltonian is a combination of the $J_1\!-\!J_2$ $XXZ$  (\ref{eqHJ1J2}), the pseudo-dipolar from  (\ref{eqH_all}),
and the field term $-h_z\sum_i S^{z_0}_i$,
where $h_z=g_z\mu_{\rm B}\mu_0 H_z$ and $g_z$ is the $z$-component of the anisotropic $g$-tensor.
For the co-aligned spins, they lead to a SWT Hamiltonian in a standard form (\ref{eq_H2})
with the parameters
\vskip -0.15cm \noindent
\begin{eqnarray}
&&\tilde{A}_{\bf k}= \frac{h_z}{3J_1S}-2\Delta\left(1+\alpha\right)+2\gamma_{{\bf k}}+2\alpha\gamma_{{\bf k}}^{(2)}, 
\nonumber\\
&&\tilde{B}_{\bf k}=\frac{4\eta}{3}\sum_\alpha e^{-i\tilde{\varphi}_\alpha} \cos k_\alpha, \ 
\label{eq_ABk_sat}
\end{eqnarray}
\vskip -0.15cm \noindent
where $k_\alpha\!=\!{\bf k}\cdot{\bm\delta}_\alpha$ and $\omega_{\bf k}\!=\!\sqrt{\tilde{A}_{\bf k}^2-\tilde{B}_{\bf k}^2}$ 
as before.

\begin{figure}[t]
\includegraphics[width=0.9\linewidth]{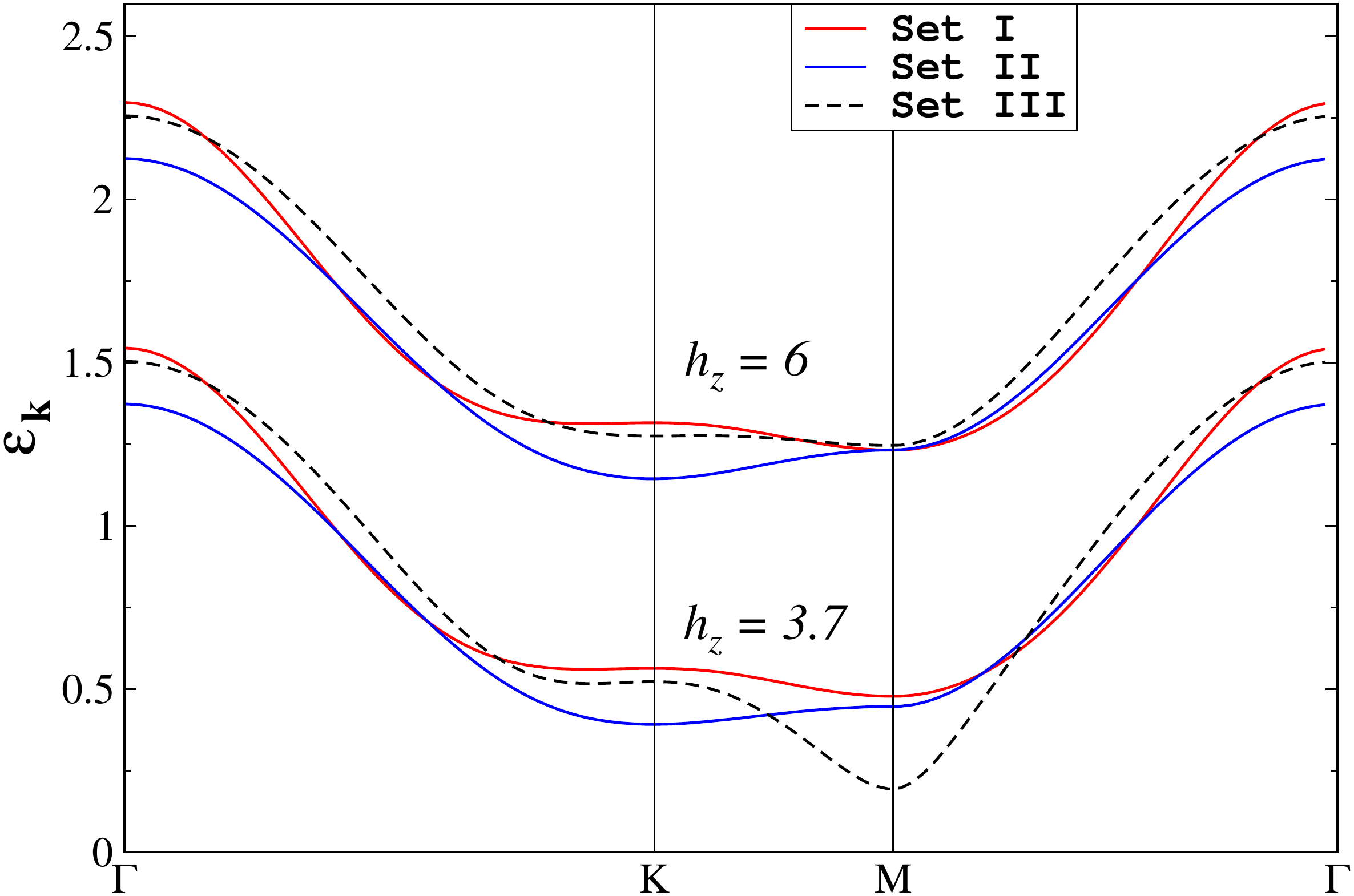}
\vskip -0.2cm
\caption{$\varepsilon_{\bf k}$ for YbMgGaO$_4$ ($3J_1S\!=\!0.327$ meV) 
in the spin-polarized phase for two representative 
fields, $h_z/3J_1 S=6$ and $3.7$, and {\tt Sets I}, {\tt II}, and {\tt III} \cite{sMM,sChen3}.
}
\label{Fig_wk_sat}
\vskip -0.5cm
\end{figure}

An important difference  of the considered case
from the more conventional models is that  although the spins are co-aligned, the 
fluctuations are not zero even in the polarized state ($\tilde{B}_{\bf k}\!\neq\!0$).
Since fluctuations are only 
due to pseudo-dipolar terms, the latter can be determined, e.~g., from the field-dependent behavior 
of the magnon dispersion. In the absence of fluctuations, high-field magnon dispersion would simply shift with the field. 
In Fig.~\ref{Fig_wk_sat} we present magnon dispersions for the two 
values of $h_z$ and for the three sets of parameters discussed above. 

The field of the transition from the stripe to the 
spin-polarized state also depends on the pseudo-dipolar terms.  
Since  the gap at the M point vanishes at the transition, 
one finds $h_s\!=\!6J_1 S(\Delta+1/3)(1+\alpha)+8|J_{\pm\pm}|S$.
Since the magnetization is not fully saturated at $H\!>\!H_s$,  
it may be possible to extract $|J_{\pm\pm}|$ from its field dependence. Experiments in YbMgGaO$_4$ 
show $M$ vs $H$ that is surprisingly linear for $H\!<\!H_s$ 
without any clear cusp indicative of a transition. The lack of the upward curvature in $M(H)$ hints at the low 
role of quantum fluctuations, characteristic of the gapped phases.

\emph{Integrated intensity of $\mathcal{S}(\mathbf{q},\omega)$.---}%
Yet another measurable quantity is the $\omega$-integrated dynamical structure factor ${\cal S}({\bf q},\omega)$.
In a fully polarized state, which is in a way identical to a ferromagnetic state, integration over the sharply-defined 
single-magnon sector yields a function that is independent of ${\bf q}$. This is precisely because of the
absence of   quantum fluctuations in the fully saturated phase. In our case,  the fluctuations are present and
${\cal S}({\bf q})$ will be modulated in ${\bf q}$. The modulation is directly proportional 
to the strength of the pseudo-dipolar terms.
While this modulation may not be easily detectable for small $J_{\pm\pm}$, such as in the {\tt Set I}, 
the lack of a significant modulation can help ruling out larger values of the pseudo-dipolar terms.

The in-plane component of the  structure factor can be obtained as
\vskip -0.15cm \noindent
\begin{equation}
\mathcal{S}(\mathbf{q}) =\frac{S}{2}\left(\frac{\tilde{A}_\mathbf{q}}{\omega_\mathbf{q}}  +
\frac{(q_y^2-q_x^2)\text{Re}\tilde{B}_\mathbf{q} -2q_x q_y\text{Im}\tilde{B}_\mathbf{q}}{q^2 \omega_\mathbf{q}}\right) ,
\label{eq_ff_2}
\end{equation}
\vskip -0.15cm \noindent
with $\tilde{A}_\mathbf{q}$ and $\tilde{B}_\mathbf{q}$ from (\ref{eq_ABk_sat}).

\begin{figure}[t]
\includegraphics[width=0.9\linewidth]{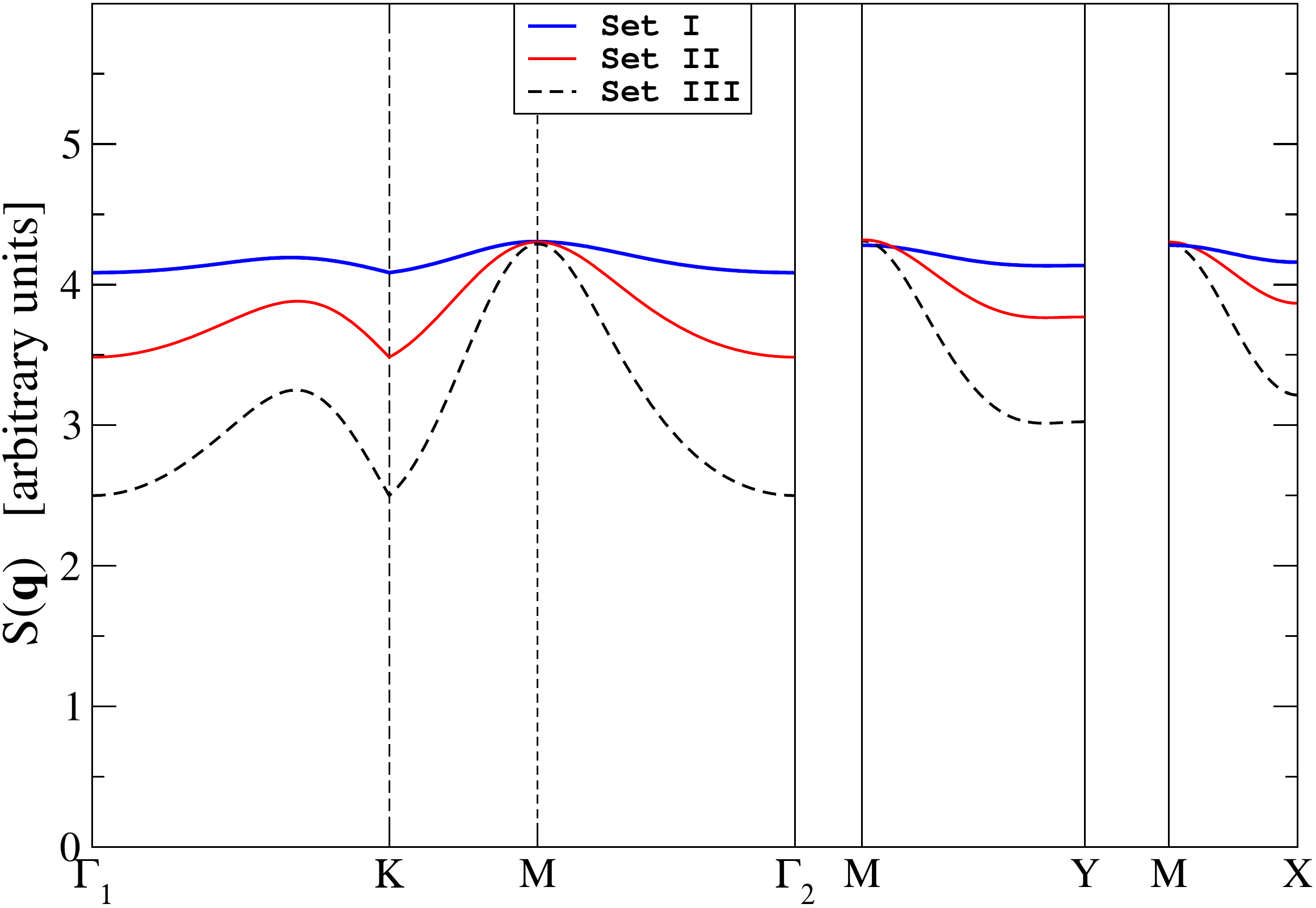}
\vskip -0.2cm
\caption{${\cal S}({\bf q})$ for the same {\tt Sets} as in Fig.~\ref{Fig_wk_sat}, $h_z/3J_1S\!=\!5.2$.}
\label{Sq1D}
\vskip -0.5cm
\end{figure}

Our Fig.~\ref{Sq1D} shows the ${\bf q}$-modulation of ${\cal S}({\bf q})$ for the three sets of parameters discussed above.
It is important to note that the momentum dependence of the $\omega$-integrated structure factor  
${\cal S}({\bf q})$ is different in the first and the subsequent Brillouin zones because of the 
explicit ${\bf q}$-dependence in the fluctuation-induced terms in (\ref{eq_ff_2}). This observation offers a yet 
another potential avenue of determining the values of the pseudo-dipolar terms.

\vspace{-0.5cm}
\subsection{Details of the DMRG calculations}
\vskip -0.3cm

For the DMRG calculations in the $6\times36$ cylinders, we perform 24 sweeps and keep up to $m=2000$ states
with truncation error less than $10^{-5}$. For the $6\times12$ cylinders, 
we perform 32 sweeps and keep up to $m=2000$ states with truncation errors less
than $10^{-6}$.
In the real-space images of cylinders, the size of the arrows represent the measurement of local spin 
with the directions of the spins in the $xy$ plane. The width of the bond on the lattice represents the 
nearest-neighbor spin-spin correlation, with ferromagnetic correlation shown as dashed and antiferromagnetic ones
as solid lines.

\begin{figure}[t]
\includegraphics[width=1\linewidth]{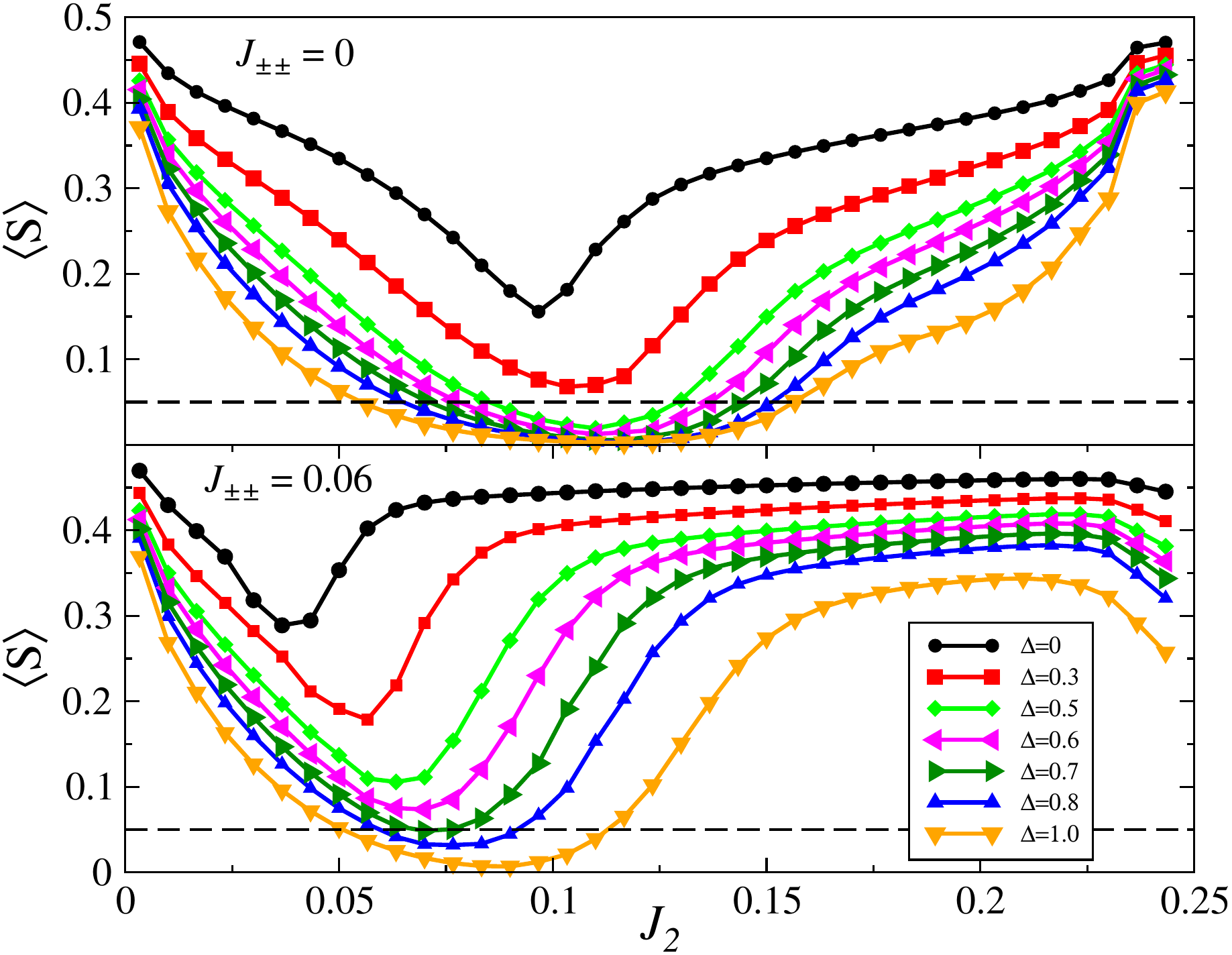}
\vskip -0.2cm
\caption{The long-cylinder DMRG scans of $\langle S\rangle$ vs $J_2$ 
for various $\Delta$'s and $J_{\pm\pm}=0$ and $J_{\pm\pm}=0.06J_1$.}
\label{S_DMRG_scans}
\vskip -0.5cm
\end{figure}

In Fig.~\ref{S_DMRG_scans}, we provide a more detailed exposition of the  long-cylinder DMRG ``scans'' of the 
$XXZ$ model (\ref{eqHJ1J2}) and of the same model with the $J_{\pm\pm}$ terms from (\ref{eqH_all}). 
The calculations are done on  $6\times36$ cylinders at fixed $\Delta$'s and $J_{\pm\pm}$. The $J_2$ is 
varied between 0 and $0.25J_1$ along the length of the cylinder.
The orders, which are verified to exist at the limiting $J_2$ values, are pinned at the boundaries.
The boundaries of the long cylinders are at $J_2=0$, where the stability of the  $120^{\degree}$ state is 
well established, or at a rather large $J_2$ where the stripe state is also well 
known to be stable. We have also performed ``narrowing'' of the window of the scan 
(range of $J_2$), to reduce the gradient of its change in order to verify the 
stability of our results and the lack of induced-order effect.
Fig.~\ref{S_DMRG_scans} shows the profiles of the ordered moments $\langle S\rangle$ vs $J_2$ for several
$\Delta$'s and for $J_{\pm\pm}\!=\!0$ and $J_{\pm\pm}\!=\!0.06J_1$. We estimated that in our clusters,
$\langle S\rangle\!=\!0.05$ line should 
separate the cases of a direct transition between the  $120^{\degree}$ and the stripe phase   
from the ones where it goes through an intermediate non-magnetic state.
This yields a criterion for the spin-liquid (SL) boundaries and it matches such boundaries  for the Heisenberg
$J_1\!-\!J_2$ model \cite{sZhuWhite}. 

We use  scans in Fig.~\ref{S_DMRG_scans} to provide  estimates for the SL boundaries, 
with  more careful verifications usually conducted on shorter $6\times12$ cylinders
for fixed values of $J_2$ via correlation functions and by studying decay of spin correlations away from the 
edges with induced orders \cite{sZhuWhite}. For instance, several points within the ``spin-liquid domes'' 
of the phase diagram in Fig.~1 of the main text have been checked to verify that the state is indeed magnetically 
disordered and that the correlation length of any induced order, either by pinning a spin with a 
field in the center or at the boundaries, falls off exponentially.

One can see for the $J_{\pm\pm}=0.06J_1$ case, that the order is strengthened and the non-magnetic region shrinks 
considerably. We estimate that $J_{\pm\pm}\approx 0.1J_1$ is sufficient to eliminate the SL state completely.

\emph{Disorder.---}%
As noted above, according to the parametrization introduced in (\ref{Jparam}), $J_{\pm\pm}\!=\!(J_{xx}-J_{yy})/4$,
the variations of the diagonal elements $J_{xx}$ and $J_{yy}$ of the exchange matrix in (\ref{H12a}) by 
$\delta J_1$ translate into variations of $J_{\pm\pm}$ by $\delta J_1/2$. 
If $J_{xx}\!\approx\! J_{yy}$, which is the case of YbMgGaO$_4$, the bare value of $J_{\pm\pm}$ is small and 
variations of the exchange matrix of order 20\%  suggested in \cite{sYuesheng17}, 
imply a spatial distribution of  completely random pseudo-dipolar $J_{\pm\pm}$ bonds of different sign.

\begin{figure}[t]
\includegraphics[width=0.7\linewidth]{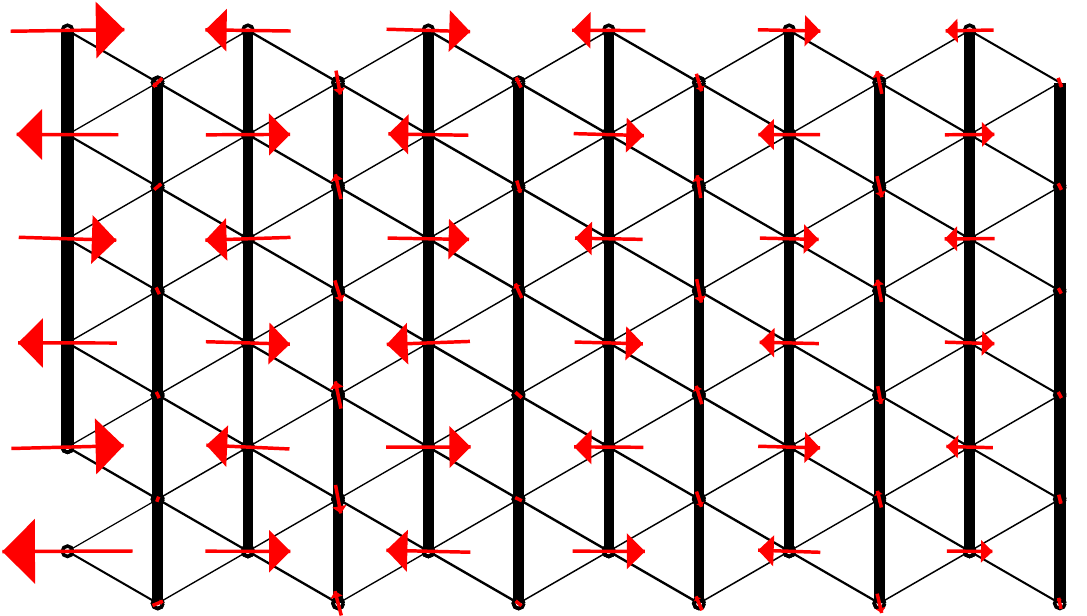}\\
\includegraphics[width=0.7\linewidth]{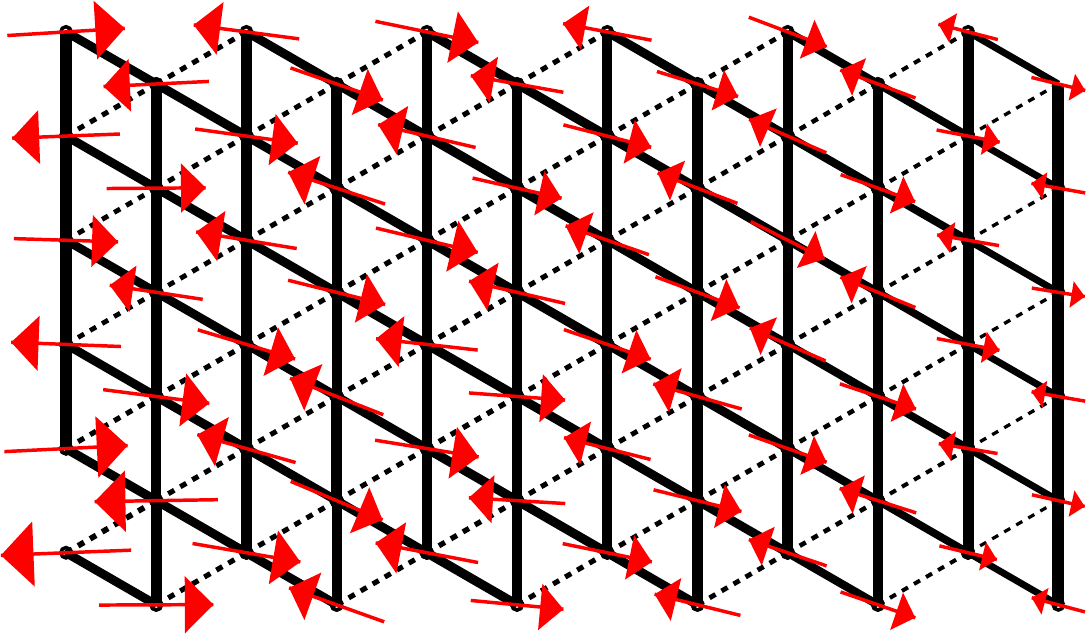}
\caption{The $6\times 12$ DMRG cylinder, $J_2=0.22J_1$, $\Delta=0.58$, and random $J_{\pm\pm}=\pm 0.05J_1$ 
with one [upper image] or two [lower image] spins at the lower left edge pinned by the field in the horizontal direction.}
\label{pinning1}
\end{figure}

\begin{figure}[t]
\includegraphics[width=0.7\linewidth]{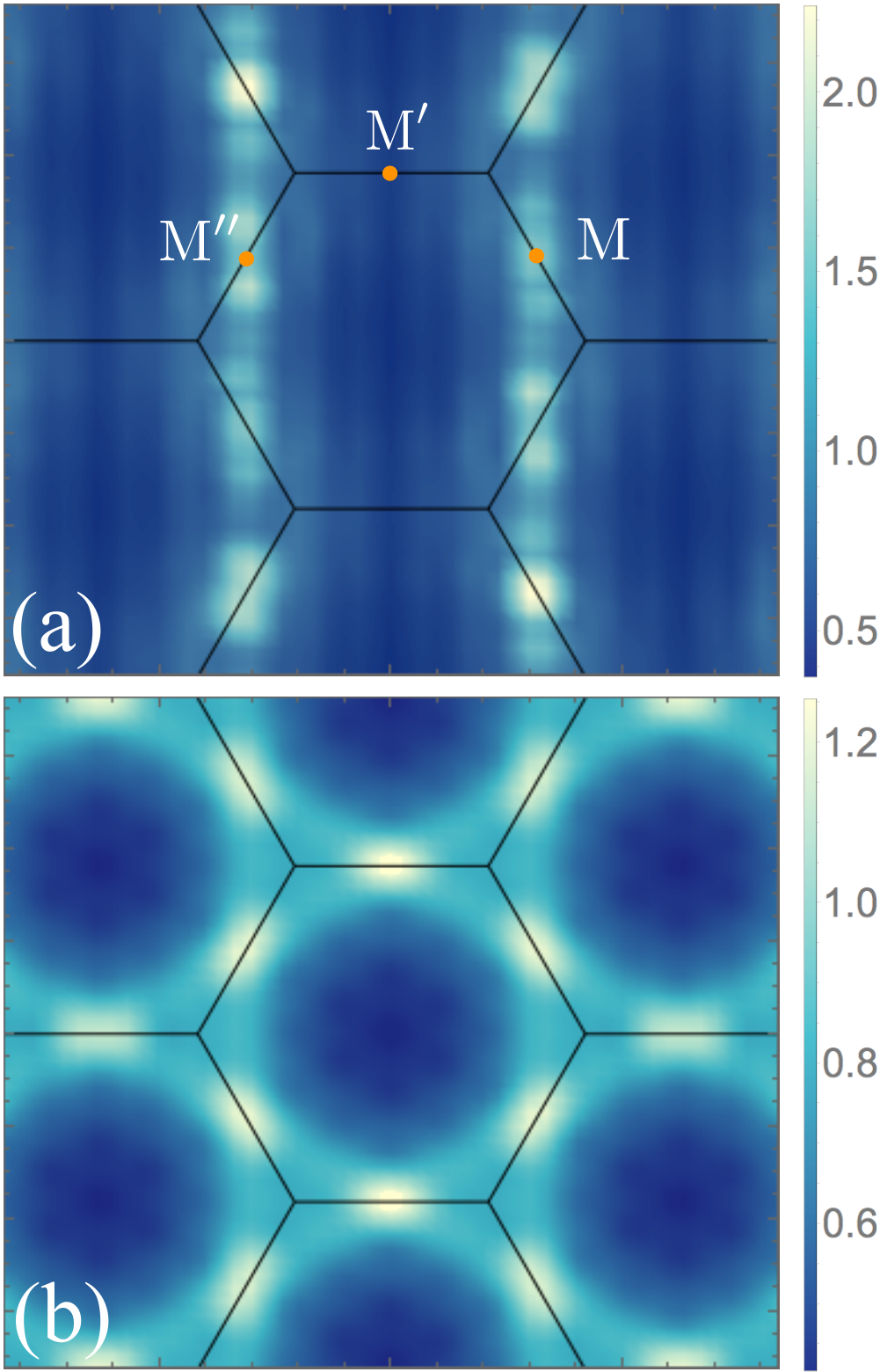}
\caption{(a) Structure factor $S(\bf q)$ for a domain-like 
disordered state for random $J_{\pm\pm}\!=\!\pm0.2J_1$. (b) Same, averaged to restore the suppressed stripe direction.}
\label{SqJpm02}
\end{figure}

In our DMRG simulations of the $XXZ$ model we use
binary distributions of $J_{\pm\pm}$ of alternating sign on a $6\times 12$ clusters.
All presented results are from individual disorder realizations. They are all 
relatively time consuming to run and to extract spin correlations. 
The final figures for the static structure factor (Fig.~4(d) of the main text and Fig.~\ref{SqJpm02}(b) below) 
included the averaging over orientations to help restore the full orientational symmetry 
of the lattice that was affected by the boundaries.

We find that for smaller values of $|J_{\pm\pm}|\!=\!0.05J_1$ and $0.1J_1$, not only an effective $U(1)$ symmetry
of the $XXZ$ model is recovered (real-space order is not pinned to the lattice), but also the $Z_3$ lattice symmetry, 
broken in each individual stripe state, is effectively restored due to a randomization of $J_{\pm\pm}$.
This is demonstrated by the structure factor  showing broadened peaks at two different M-points 
associated with two different stripe orderings, with the third stripe direction strongly suppressed by the cluster geometry,
see main text.
We refer to these states as to the \emph{stripe-superposition} states, in which  stripe orders coexist in a fluctuating
manner. 

The details of this unusual state are revealed by probing it with various pinning fields. 
In the first one, one spin is pinned at the edge (lower left corner), see Fig.~\ref{pinning1}.
The resulting pattern over the whole cluster is associated with an almost equal superposition of the two 
types of stripes, with M and M$^{\prime\prime}$ ordering vectors,  running across each other. 
In this case, the structure factor
continues to show two peaks at these two M-points even with the strong pinning field. 
In the second such experiment, we apply a pinning field to a second site in the next column in order to favor 
one of the stripe orders, see Fig.~\ref{pinning1}. 
That state is indeed revealed, the single-stripe real-space ordered  pattern emerges, 
and the second peak in $S(\bf q)$ becomes suppressed when the pinning field is made strong.  

For smaller $J_{\pm\pm}$ and for most disorder realizations, we found 
stripe-superposition states, and for some other realizations the state was a robust 
stripe state of one orientation. This is consistent with most disorder distributions 
allowing fluctuations between the stripes to exist within the domains
of the size of our clusters, and for the other distributions, which for some 
reason are more biased, disorder is pinning one stripe orientation domain.

For larger values of the randomized $J_{\pm\pm}\!=\!\pm0.2J_1$, we found disordered spin 
domains with mixed stripe orientations forming static structures with large ordered
moments, see main text. Here we show the associated structure factor, see Fig.~\ref{SqJpm02}.
For larger random $J_{\pm\pm}\!=\!\pm0.2$, the correlation length becomes compatible with
the cluster size, and we have observed either robust stripes or a coexistence 
of the stripe domains within a cluster, as shown in our Fig.~4(b) of the main text.  
The shorter stripe domain sizes at stronger disorder (correlation length $\sim 6$
lattice sites for $J_{\pm\pm}\!=\!\pm0.2$) is consistent with a general expectation.

\begin{figure}[t]
\includegraphics[width=0.9\linewidth]{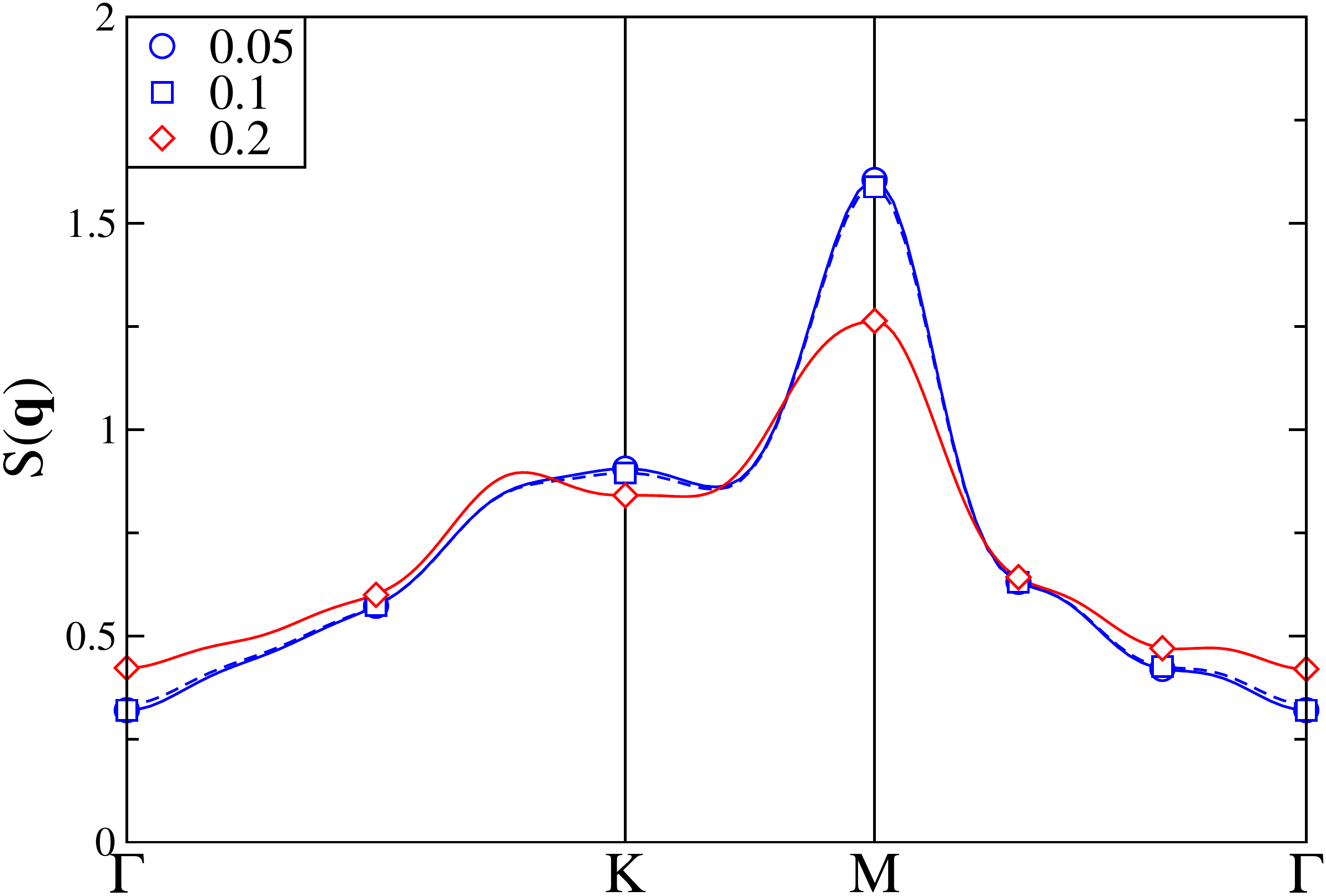}
\caption{1D cuts of the averaged structure factor $S(\bf q)$.}
\label{1Dcuts}
\end{figure}

In the last Figure \ref{1Dcuts}, we compare  1D cuts of the 
orientationally-averaged structure factor $S(\bf q)$ for the 
stripe-superposition states for smaller random  $J_{\pm\pm}\!=\!\pm0.05J_1$ and $\pm0.1J_1$ 
from the main text with the glass-like $S(\bf q)$ for larger $J_{\pm\pm}\!=\!\pm0.2J_1$ in Fig.~\ref{SqJpm02}.
Both cases, the glass-like mixed-stripe and the stripe-superposition states, result in the $S(\bf q)$ structure that
is strongly reminiscent of the experimental results in YbMgGaO$_4$ \cite{sMM}.

\emph{Note on the $J_{z\pm}$ terms.---}%
We have performed some preliminary DMRG calculations \cite{sZhu} of the $J_1$-only $XXZ$ model with the 
so-called $J_{z\pm}$ terms \cite{sChen1,sChen2}, see (\ref{eqH_1b}), 
which were initially dropped in the context of YbMgGaO$_4$ as they were found to be small \cite{sChen1}.
These calculations show a direct transition between robust $120^{\degree}$ and stripe phases at rather large 
$J_{z\pm}\approx 0.3 J_1$ with no indication of a spin-liquid away from the Heisenberg limit of the  $XXZ$ term.

\vspace{-0.3cm}


\end{document}